\documentclass[12pt,a4paper,final]{iopart}
\usepackage[utf8]{inputenc}
\usepackage{amsmath,amssymb,amsfonts}
\usepackage{cite}
\usepackage[breaklinks=true,colorlinks=true,linkcolor=blue,urlcolor=blue,citecolor=blue]{hyperref}
\hypersetup{breaklinks=true,colorlinks=true,citecolor=violet,urlcolor=blue!60!black,linkcolor=red}
\usepackage{braket}
\usepackage{pgf,tikz}
\usetikzlibrary{decorations.markings}
\tikzset{>=latex} 
\usetikzlibrary{shapes,calc,decorations.markings}
\usepackage{empheq}
\usepackage[most]{tcolorbox}
\usepackage{verbatim}

\newtcbox{\othermathbox}[1][]{nobeforeafter, math upper, tcbox raise base, 
          enhanced, sharp corners, colback=black!5, colframe=black, 
           left=0.7em, top=0.5em, right=0.7em, bottom=0.5em}
           
\newcommand{\ci}{\mathrm{i}\mkern1mu}
\newcommand{\tH}{\mathtt{h}}
\definecolor{mycolor}{rgb}{0,0,0.5}

\begin{document}

\title{Exact time evolution formulae in the XXZ spin chain with domain wall initial state}
\author{Jean-Marie St\'ephan}
\address{
Univ Lyon, Universit\'e Claude Bernard Lyon 1, CNRS UMR 5208, Institut Camille Jordan, 43 blvd. du 11 novembre 1918, F-69622 Villeurbanne cedex, France
}
\date{\today}

\begin{abstract}
We study the time evolution of the spin-1/2 XXZ chain initialized in a domain wall state, where all spins to the left of the origin are up, all spins to its right are down. The focus is on exact formulae, which hold for arbitrary finite (real or imaginary) time. In particular, we compute the amplitudes corresponding to the process where all but $k$ spins come back to their initial orientation, as a $k-$fold contour integral. These results are obtained using a correspondence with the six vertex model, and taking a somewhat complicated Hamiltonian/Trotter-type limit. Several simple applications are studied and also discussed in a broader context.   
\end{abstract}

\maketitle
\tableofcontents
\newpage

\section{Introduction and main results}
\label{sec:introduction}
\subsection{Context}
\label{sec:context}
Since the seminal work of Bethe \cite{Bethe}, a considerable amount of energy has been invested in exploiting his Ansatz to extract meaningful physical information in what are now called quantum integrable systems. Some of those results deal with thermodynamics at finite temperature, algebraic tools \cite{korepin_bogoliubov_izergin_1993} to study static or dynamical correlations, asymptotic results for those and their relation to conformal field theories, relation to statistical mechanics through the six vertex model \cite{Baxter1982}, stochastic processes, just to name a few. 

A fairly recent trend is to investigate inhomogeneous versions of such $1d$ quantum integrable models. This is relevant experimentally, as such models may be realized in cold atomic systems, where the presence of a trapping potential is almost unavoidable\cite{Cazalilla}. Another huge class of problems which lead to inhomogenous situations goes under the name quantum quench, where a quantum system is prepared in a given (perhaps very simple) state, and let evolve unitarily with the integrable Hamiltonian. While many works have studied thermalization (or lack thereof \cite{GGE, Cradle}) in such out of equilibrium setups, one can also consider inhomogeneous initial states which lead to non-trivial time dependent behavior. Inhomogenous systems also occur on the statistical mechanical side, as exemplified by the celebrated arctic phenomenon for dimer or vertex models \cite{jockusch1998random,Abanov_hydro,kenyon2007,ColomoPronkocurve,ColomoPronkoZinn,deGier2021,Stephan_extreme,Aggarwal2020}, where non-trivial density profiles may emerge in the scaling limit. 

A straightforward strategy to study such questions is to use the (algebraic or not) Bethe Ansatz, and try to write down explicit expressions in finite size/finite time for observables of interest, in or out of equilibrium. One can then, in principle, obtain the relevant physical results by taking a scaling limit with large system size and/or large time, in which case new and sometimes universal phenomena may arise. In the special case of free fermions this can be done rather systematically. One typically obtains integral formulae for the two-point function, with an explicit integrand. A saddle point treatment of those integrals then leads to considerable simplifications. These saddle point equations have a simple hydrodynamic interpretation, where typically a quasiparticle with momentum $k$ simply propagates at a known  group velocity $v(k)$. This means a fundamentally hydrodynamic treatment sidesteps this asymptotic analysis, and is very often the simplest way to obtain exact results in the scaling limit. A hydrodynamic solution to arctic problems can also be found for free fermions, see  Refs~\cite{Abanov_hydro,kenyon2007}. In this case limiting density profiles are often less explicit, but nevertheless parametrized by analytic functions. 

An obvious question is whether hydrodynamics can also be applied to \emph{interacting} integrable systems out of equilibrium. The answer is yes, as was shown recently by two independent groups \cite{YoungItalians,Doyonhydro} (see \cite{Alba_2021,Bastianello_2021,bouchoule2021generalized} for state of the art reviews). The tricky part compared to standard hydrodynamics is to incorporate the extensive number of conserved quantities which are the trademark of integrable systems. Hence the name \emph{generalized hydrodynamics} (GHD). Besides standard hydrodynamic assumptions such as separation of space time scales and local equilibration, a key technical  requirement is the so-called root condensation hypothesis, which tells us that the discrete set of Bethe solutions called rapidities (generalizing momenta in the free case) becomes a non-trivial but manageable "thermodynamic Bethe Ansatz" (TBA) density in the thermodynamic limit. The GHD equation is then a purely classical time evolution equation for the TBA content of the initial state, similar to an Euler equation in the case of conventional hydrodynamics. Several relevant  initial states have a well understood TBA content, which means one can (usually numerically) solve the hydrodynamic equations, and determine non-trivial local features of the time evolved state of the system.  Of course, subleading effects can be investigated also \cite{DeNardis}, including quantum effects  \cite{QGHD,Fagotti_higher}. 

One can try to apply similar arguments on the statistical mechanical side, where time evolution is now imaginary, but there are important differences compared to real time. First, the infinite number of conserved quantities plays less of a physical role, since those are not conserved by the imaginary time evolution. However, the 
standard statistical mechanical framework based on free energy minimisation holds irrespective of (local) interactions, meaning there is no fundamental obstruction to writing hydrodynamic equations for that class of problems too. The devil unfortunately lies in the technical details. Understanding the TBA behind this yields somewhat non standard problems with large imaginary twists\cite{Reshetikhin_lectures,Granet_2019,Granet2_2019}, for which root condensation is expected to occur, but exact calculations of the root densities becomes very difficult (see however \cite{deGier2021}). This is in part due to the fact that roots condense on non-trivial curves in the complex plane, and those are not so easy to determine (in contrasts with the famous string hypothesis in the absence of a twist, which postulates that the relevant complex roots lie on straight lines \cite{takahashi_1999} with constant imaginary part).  

From all the above considerations it is tempting to ask whether a derivation of such (real time) hydrodynamic principles can be performed starting from the lattice also in the presence of interactions. While some parts of GHD can be put on firm grounds (see e.g. \cite{Pozsgay_almostproof, Pozsgay_almostproof2,Cubero_microscopic_2021}), this is known to be a very difficult if not hopeless task. Less ambitiously, one could ask for \emph{one example} of an interacting quench for which exact lattice calculations may be compared to GHD predictions, but no such example is known. This last question motivates the present study.  We will consider a known and simple to formulate interacting problem which has both real time \cite{antal1999transport,Gobert,antal_2008,YoungItalians} and imaginary time \cite{prahofer2002scale,Allegra_2016} interpretations. The GHD equation can even be solved explicitly in that case \cite{ColluraDeLucaViti}. Extending the exact result for the return amplitude \cite{Stephan_return}, we write down exact finite time formulae for more complicated amplitudes, and discuss how some very simple physical information can extracted out of those. While the results demonstrated here are just one step in this direction and clearly do not allow to make contact with hydrodynamics yet, it is our hope that they will be useful in better understanding the emergence of GHD for this quench protocol, or find a hydrodynamic solution to the imaginary time problem, which is still not known for the technical reasons explained above. Most of our examples will deal with imaginary time, but we will regularly comment on the analogous real time problem.

\subsection{Problem studied and exact formulae}
\label{sec:theproblem}
In this paper, we consider the Hamiltonian of the spin-1/2 XXZ chain on the infinite line
\begin{equation}\label{eq:H}
    H=\sum_{j\in \mathbb{Z}}\left[ \sigma_j^+\sigma_{j+1}^-+\sigma_j^-\sigma_{j+1}^++\frac{\Delta}{2}\left(\sigma_j^z\sigma_{j+1}^z-1\right)\right],
\end{equation}
where the $\sigma_j^\alpha$ are standard Pauli matrices at site $j$ \cite{korepin_bogoliubov_izergin_1993}, 
and $\Delta$ is the anisotropy parameter. As usual, spins are measured in the basis generated by the eigenstates of the $\sigma_j^z$. A special role will be played by the inhomogeneous initial state
\begin{equation}
    \ket{\psi_0}=\ket{\ldots\uparrow\uparrow\uparrow\uparrow\downarrow\downarrow\downarrow\downarrow\ldots}
\end{equation}
which has all spins up for $j\leq 0$, and all spins down for $j>0$. One can also interpret up spins as particles and down spins as holes, in which case all sites to the left of the origin are occupied, all sites to its right are empty. In the following, $\ket{\psi_0}$ will serve as a reference state, out of which all other states may be constructed. The partition function
\begin{equation}\label{eq:Z}
    Z(\tau)=\braket{\psi_0|e^{\tau H}|\psi_0}
\end{equation}
is finite and was computed in Ref.~\cite{Stephan_return}, as a Fredholm determinant of an operator with an explicit kernel which we recall later on. The formula holds for arbitrary values of the anisotropy parameter $\Delta\in\mathbb{R}$, arbitrary $\tau\in\mathbb{C}$ --and so will all other formulae discussed in this introduction. In real time (imaginary $\tau$) this is a return amplitude after a quench from the initial state $\ket{\psi_0}$, while in imaginary time (real $\tau$) the return amplitude has an obvious statistical mechanical interpretation. 

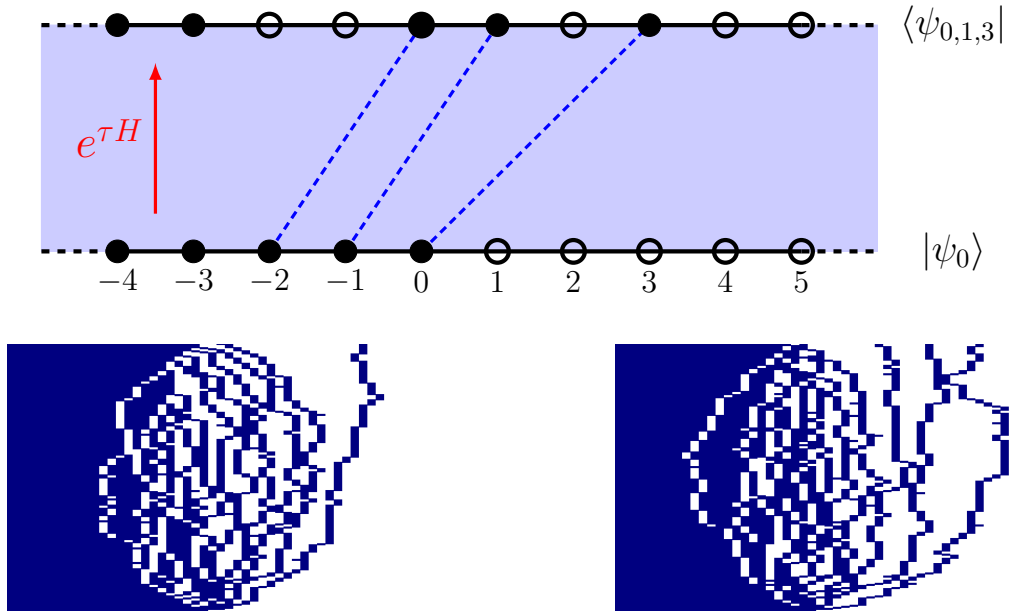
\begin{figure}[htbp]
\centering
\begin{tikzpicture}
\filldraw[blue,opacity=0.2] (-1,0) -- (10,0) -- (10,3) -- (-1,3) -- cycle;
\draw[very thick,->,red] (0.5,0.5) -- (0.5,2.5);
\draw[red] (-0.1,1.5) node {\Large{$e^{\tau H}$}};
\draw (11,0) node {\large{$\ket{\psi_0}$}};
\draw (11,3) node {\large{$\bra{\psi_{0,1,3}}$}};
\draw[ultra thick,dashed] (-1,0) -- (0,0);
\draw[ultra thick,dashed] (-1,3) -- (0,3);
\draw[ultra thick,dashed] (9,0) -- (10,0);
\draw[ultra thick,dashed] (9,3) -- (10,3);
\draw[ultra thick] (0,0) -- (9,0);
\draw[ultra thick] (0,3) -- (9,3);
\draw (0,-0.4) node {$-4$};\
\draw (1,-0.4) node {$-3$};\draw (2,-0.4) node {$-2$};
\draw (3,-0.4) node {$-1$};\draw (4,-0.4) node {$0$};
\draw (5,-0.4) node {$1$};\draw (6,-0.4) node {$2$};
\draw (7,-0.4) node {$3$};\draw (8,-0.4) node {$4$};
\draw (9,-0.4) node {$5$};
\draw[densely dashed,blue,very thick] (2,0) -- (4,3);
\draw[densely dashed,blue,very thick] (3,0) -- (5,3);
\draw[densely dashed,blue,very thick] (4,0) -- (7,3);
\foreach \x in {0,1,2,3,4}{
\filldraw (\x,0) circle (0.15cm);
\draw[ultra thick] (\x+5,0) circle (0.15cm);
}
\foreach \x in {0,1,4}{
\filldraw (\x,3) circle (0.15cm);
}
\draw[ultra thick] (2,3) circle (0.15cm);
\draw[ultra thick] (3,3) circle (0.15cm);
\draw[ultra thick] (4,3) circle (0.15cm);
\filldraw (5,3) circle (0.15cm);
\draw[ultra thick] (6,3) circle (0.15cm);
\filldraw (7,3) circle (0.15cm);
\draw[ultra thick] (8,3) circle (0.15cm);
\draw[ultra thick] (9,3) circle (0.15cm);
\begin{scope}[yshift=-3.2cm,xshift=1.3cm]
 \node (amp1) at (0,0.2) {\scalebox{1}[-1]{\includegraphics[width=5.5cm]{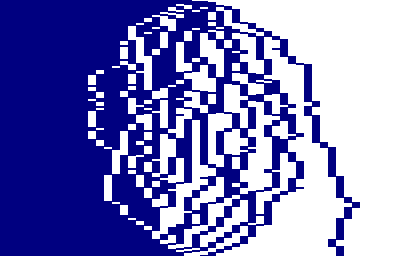}}};
  \node (amp3) at (8,0.2) {\scalebox{1}[-1]{\includegraphics[width=5.5cm]{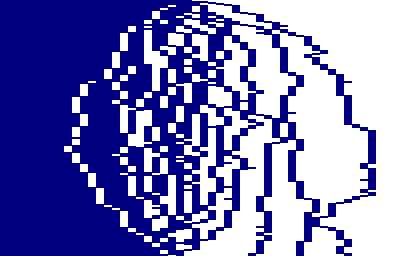}}};
\end{scope}
\end{tikzpicture}
\caption{\emph{Top:} Amplitude $A_{0,1,3}(\tau)$, corresponding to the process where particles at respective positions $-2,-1,0$ moved to positions $0,1,3$, and all the others came back to their initial locations. \emph{Bottom:} Typical configurations for $\tau=16$. Left: one particle amplitude $A_x(\tau)$. Right: three particle amplitude $A_{x_1,x_2,x_3}(\tau)$. The samples are obtained by Monte Carlo simulation of the six vertex model with a small value of $b$, using the correspondence explained in section \ref{sec:derivation}, similar to pictures in \cite{Stephan_extreme}.}
\label{fig:amplitudes}
\end{figure}

For any integers $x_1,\ldots,x_l$ that satisfy $-l+1\leq x_l<\ldots<x_1$, define the states
\begin{equation}\label{eq:psi_notation}
     \ket{\psi_{x_l,\ldots,x_1}}=\sigma_{x_l}^+\sigma_{-l+1}^-\ldots\sigma_{x_2}^+\sigma_{-1}^-\sigma_{x_1}^+\sigma_0^-\ket{\psi_0}.
\end{equation}
In words, these are obtained from $\ket{\psi_0}$ by moving the particles at positions $-l+1,\ldots,0$ somewhere to the right, at positions $x_l,\ldots,x_1$. The main goal of this paper is to study the time evolution from the domain-wall state, in particular the  amplitudes
\begin{equation}
    A_{x_l,\ldots,x_1}(\tau)=\frac{\braket{\psi_{x_l,\ldots,x_1}|e^{\tau H}|\psi_0}}{\braket{\psi_0|e^{\tau H}|\psi_0}},
\end{equation}
 shown in figure~\ref{fig:amplitudes}. $l$ refers to the number of particles which did not go back to their initial positions, but this amplitude formally already involves an infinite number of them.

To state the main results, it is necessary to first introduce the generating function
\begin{align}\label{eq:hdef}
    h(\tau|z)&=\sum_{x\geq 0} A_x(\tau) z^x
\end{align}
for the one-particle amplitude. Notice $h(0|z)=1=h(\tau|0)$. Alternatively, one can write a power series with respect to $\tau$, in which case
\begin{equation}
    h(\tau|z)=\sum_{m\geq 0}p_m(z) \tau^m,
\end{equation}
where one can check that $p_m$ is a polynomial of degree $m$, with coefficients that depend on $\Delta$ (see  \ref{app:checks}).  
Determining the first few $p_m$ is also straightforward, since the action of the first few powers of $H$ on the initial state $\ket{\psi_0}$ generates only few states. 

We will show that this generating function $h(\tau|z)$ satisfies the exact second order partial differential equation (PDE):
\begin{empheq}[box=\othermathbox]{align}\label{eq:pde}
  \left(\tau\partial_\tau^2+\left[1-2\tau\left(\frac{1}{z}-\Delta\right)\right]\partial_\tau+Q(\tau)-z+\Delta\right)h(\tau|z)=
    \left(1-2\Delta z+z^2\right)\partial_z h(\tau|z),
\end{empheq}
where we have introduced the function 
\begin{equation}\label{eq:Y}
 Q(\tau)=2\tau\frac{d^2\log Z(\tau)}{d\tau^2}+\frac{d\log Z(\tau)}{d\tau}.  
\end{equation}
In terms of $h(\tau|z)$, the amplitudes are given by the exact multiple contour integral
\begin{empheq}[box=\othermathbox]{align}\label{eq:multiamplitude}
     A_{x_l,\ldots,x_1}(\tau)= \oint_{\mathcal{C}^l} \prod_{j=1}^l \frac{dz_j}{2\ci\pi z_j^{x_j+l}}\, \frac{\displaystyle{\det_{1\leq j,k\leq l}\left(z_k^{l-j}\left[1-z_k\partial_\tau\right]^{j-1} h(\tau|z_k) \right)}}
     {\displaystyle{\prod_{1\leq j<k\leq l}(z_j z_k-2\Delta z_k+1)}},
\end{empheq}
where $\mathcal{C}$ is any (sufficiently small) counterclockwise oriented contour enclosing the origin.
Equations (\ref{eq:pde}) and (\ref{eq:multiamplitude}) are the main result of this paper. Their derivation is presented in section \ref{sec:derivation}, using a relation to the six vertex model. The reader more interested in applications may skip this section at first reading.

It is worth noting that all physical information can in principle be extracted from the partition function $Z(\tau)$, since its precise knowledge allows to determine $h(\tau|z)$ through (\ref{eq:Y}), (\ref{eq:pde}), and then $A_{x_l,\ldots,x_1}(\tau)$ through (\ref{eq:multiamplitude}). The amplitudes can then  be used to reconstruct any correlation function corresponding to real or imaginary time evolution setups with domain wall initial states. This logic will be illustrated in section~\ref{sec:examples}, on simple examples. The results are further discussed in section~\ref{sec:conclusion}, where we conclude. Some technical aspects are relegated to two appendices. In \ref{app:orthopolys} we establish a result on orthogonal polynomials which is essential to the derivation. \ref{app:checks} presents power series expansions of $Z(\tau)$ and $h(\tau|z)$ to high orders in $\tau$, which allows for very strong checks of the results.

\section{Derivation of the main formulae}
\label{sec:derivation}
Equation (\ref{eq:multiamplitude}) looks very similar to exact results which may be found in the literature on the six vertex model with domain wall boundary conditions \cite{ColomoPronkotopbottom,CantiniColomoPronko,ColomoDiGiulioPronko}. In fact, it is nothing but a limit of one such formula, even though performing this task still requires, as we shall see, a significant amount of work. The connection was already used in Ref.~\cite{Stephan_return} to compute exactly the partition function. We recall it in section  \ref{sec:relation}, before proceeding with the actual derivation in section~\ref{sec:hamlimit}.  
\begin{figure}[htbp]
\centering  
 \begin{tikzpicture}[scale=1.2]
  \draw[very thick] (0,0) -- (1,1);
  \draw[very thick] (0,1) -- (1,0);
  \draw[line width=4pt,color=mycolor] (0,0) -- (1,1);
  \draw[line width=4pt,color=mycolor] (0,1) -- (1,0);
  \draw (0.5,-0.5) node {$a_1$};
  \begin{scope}[xshift=2cm]
  \draw[thick] (0,0) -- (1,1);
  \draw[thick] (0,1) -- (1,0);
  \draw (0.5,-0.5) node {$a_2$};
  \end{scope}
    \begin{scope}[xshift=5cm]
  \draw[thick] (0,0) -- (1,1);
  \draw[thick] (0,1) -- (1,0);
  \draw[line width=4pt,color=mycolor] (0,0) -- (1,1);
  \draw (0.5,-0.5) node {$b_1$};
  \end{scope}
    \begin{scope}[xshift=7cm]
  \draw[thick] (0,0) -- (1,1);
  \draw[thick] (0,1) -- (1,0);
  \draw[line width=4pt,color=mycolor] (0,1) -- (1,0);
  \draw (0.5,-0.5) node {$b_2$};
  \end{scope}
    \begin{scope}[xshift=10cm]
  \draw[thick] (0,0) -- (1,1);
  \draw[thick] (0,1) -- (1,0);
  \draw[line width=4pt,color=mycolor] (0,0) -- (0.5,0.5);
  \draw[line width=4pt,color=mycolor] (0,1) -- (0.5,0.5);
  \draw (0.5,-0.5) node {$c_1$};
  \end{scope}
    \begin{scope}[xshift=12cm]
  \draw[thick] (0,0) -- (1,1);
  \draw[thick] (0,1) -- (1,0);
  \draw[line width=4pt,color=mycolor] (1,0) -- (0.5,0.5);
  \draw[line width=4pt,color=mycolor] (1,1) -- (0.5,0.5);
  \draw (0.5,-0.5) node {$c_2$};
  \end{scope}
 \end{tikzpicture}
\label{fig:6v_weights}
\caption{Weights of the six vertex model, seen as osculating lattice trajectories (thick blue lines). In the following we assume a 'particle-hole' symmetry $a_1=a_2=a$, $b_1=b_2=b$, $c_1=c_2=c$.}
\end{figure}

\subsection{Relation to the six vertex model}
\label{sec:relation}
The six vertex model is a well known integrable model in statistical mechanics \cite{Baxter1982}. We use here a formulation in terms of lattice paths, which can touch but never cross. The weights of the model are shown in figure \ref{fig:6v_weights}. The corresponding anisotropy parameter is defined as
\begin{equation}
    \Delta=\frac{a^2+b^2-c^2}{2ab}.
\end{equation}

\begin{figure}[htbp]
\begin{tikzpicture}[scale=1.2]
\draw (7.25,1.5) node {\begin{tcolorbox}[width=1.5cm,right=0.1cm,left=0.1cm,colframe=green!50!black,colback=green!10!white]
$b=1$
\end{tcolorbox}};
 \draw (6.5,0) node {$0$};
  \draw (6.5,3) node {$\tau$};
  \draw[thin] (0,0) -- (3,3);
  \foreach \x in {1,2,3,4,5}{
  \draw[thin] (0.5*\x,0) -- (3,3-0.5*\x);
  \draw[thin] (0,0.5*\x) -- (3-0.5*\x,3);
  }
  \draw[thin] (0,3) -- (3,0);
  \foreach \x in {1,2,3,4,5}{
  \draw[thin] (3,0.5*\x) -- (0.5*\x,3);
  \draw[thin] (0.5*\x,0) -- (0,0.5*\x);
  }
\foreach \y in {0,1,2,3,4,5}{
\draw[thick,densely dotted] (0,0.5*\y) -- (-0.25,0.5*\y+0.25);
\draw[thick,densely dotted] (0,0.5*\y+0.5) -- (-0.25,0.5*\y+0.25);
\draw[thick,densely dotted] (6,0.5*\y) -- (6.25,0.5*\y+0.25);
\draw[thick,densely dotted] (6,0.5*\y+0.5) -- (6.25,0.5*\y+0.25);
}
\foreach \x in {0,1,2,3,4,5}{
\draw[mycolor,line width=2pt] (2.75-0.5*\x,0.25) -- (0,3-0.5*\x);
\draw[mycolor,line width=2pt] (0,0.5*\x) -- (3-0.5*\x,3);
}
\foreach  \x in {0,1,2,3,4}{
\draw[mycolor,line width=2pt] (0.75+0.5*\x,0.25) -- (2+0.25*\x,1.5-0.25*\x);
\draw[mycolor,line width=2pt] (0.75+0.5*\x,2.75) -- (2+0.25*\x,1.5+0.25*\x);
}
\foreach \x in {0,1,2,3,4,5}{
\draw[black,line width=3pt] (0.5*\x,0) -- (0.25+0.5*\x,0.25);
\draw[black,line width=3pt] (0.25+0.5*\x,0.25) -- (0.5+0.5*\x,0);
\draw[black,line width=3pt] (0.5*\x,3) -- (0.25+0.5*\x,2.75);
\draw[black,line width=3pt] (0.25+0.5*\x,2.75) -- (0.5+0.5*\x,3);
}
\draw[line width=3pt,red,densely dashed] (3,0.5) -- (2,1.5) -- (3,2.5) -- (4,1.5) -- cycle;
\begin{scope}[xshift=3cm]
   \draw (0,0) -- (3,3);
  \foreach \x in {1,2,3,4,5}{
  \draw (0.5*\x,0) -- (3,3-0.5*\x);
  \draw (0,0.5*\x) -- (3-0.5*\x,3);
  }
  \draw (0,3) -- (3,0);
  \foreach \x in {1,2,3,4,5}{
  \draw (3,0.5*\x) -- (0.5*\x,3);
  \draw (0.5*\x,0) -- (0,0.5*\x);
  }
\foreach \x in {0,1,2}{
}
\end{scope}
 \node (ham1) at (10.5,1.5) {\scalebox{1}[-1]{\includegraphics[width=5.5cm]{ 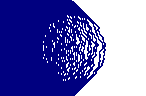}}};
 
\begin{scope}[yshift=-3.5cm]
\draw (7.25,1.5) node {\begin{tcolorbox}[width=1.5cm,right=0.1cm,left=0.1cm,colframe=green!50!black,colback=green!10!white]
$b=\frac{1}{2}$
\end{tcolorbox}};
 \draw (6.5,0) node {$0$};
  \draw (6.5,3) node {$\tau$};
 \draw (0,0) -- (6,3);
 \foreach \x in {1,2,3,4,5,6,7,8,9,10,11}{
  \draw (0.5*\x,0) -- (6,3-0.25*\x);
  \draw (0,0.25*\x) -- (6-0.5*\x,3);
  }
  \draw (0,3) -- (6,0);
  \foreach \x in {1,2,3,4,5,6,7,8,9,10,11}{
  \draw (6,0.25*\x) -- (0.5*\x,3);
  \draw (0.5*\x,0) -- (0,0.25*\x);
  }

  \foreach \x in {0,1,2,3,4,5}{
\draw[color=mycolor,line width=2pt] (2.75-0.5*\x,0.125) -- (0,1.5-0.25*\x);
\draw[color=mycolor,line width=2pt] (0,1.5+0.25*\x) -- (3-0.5*\x,3);
}
\foreach  \x in {0,1,2,3,4,5}{
\draw[color=mycolor,line width=2pt] (0.25+0.5*\x,0.125) -- (1.75+0.25*\x,0.875-0.125*\x);
\draw[color=mycolor,line width=2pt] (0.25+0.5*\x,2.875) -- (1.75+0.25*\x,2.125+0.125*\x);
}
\foreach \x in {1,2,3,4,5}{
\draw[color=mycolor,line width=2pt] (0,0.25*\x) -- (1.75-0.25*\x,0.875+0.125*\x);
\draw[color=mycolor,line width=2pt] (0,3-0.25*\x) -- (1.75-0.25*\x,2.125-0.125*\x);
}
\foreach \y in {0,1,2,3,4,5,6,7,8,9,10,11}{
\draw[thick,densely dotted] (0,0.25*\y) -- (-0.25,0.25*\y+0.125);
\draw[thick,densely dotted] (0,0.25*\y+0.25) -- (-0.25,0.25*\y+0.125);
\draw[thick,densely dotted] (6,0.25*\y) -- (6.25,0.25*\y+0.125);
\draw[thick,densely dotted] (6,0.25*\y+0.25) -- (6.25,0.25*\y+0.125);
}

  \foreach \x in {0,1,2,3,4,5}{
\draw[black,line width=3pt] (0.5*\x,0) -- (0.25+0.5*\x,0.125);
\draw[black,line width=3pt] (0.25+0.5*\x,0.125) -- (0.5+0.5*\x,0);
\draw[black,line width=3pt] (0.5*\x,3) -- (0.25+0.5*\x,2.875);
\draw[black,line width=3pt] (0.25+0.5*\x,2.875) -- (0.5+0.5*\x,3);
}
\draw[line width=3pt,red,densely dashed] (3,0.25) -- (0.5,1.5) -- (3,2.75) -- (5.5,1.5) -- cycle;
 \node (ham2) at (10.5,1.5) {\scalebox{1}[-1]{\includegraphics[width=5.5cm]{ 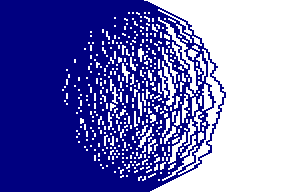}}};
\end{scope}
\begin{scope}[yshift=-7cm]
\draw (7.25,1.5) node {\begin{tcolorbox}[width=1.6cm,right=0.1cm,left=0.1cm,colframe=green!50!black,colback=green!10!white]
$b\to 0$
\end{tcolorbox}};
 \draw (6.5,0) node {$0$};
  \draw (6.5,3) node {$\tau$};
\fill[color=blue!20] (0,0) -- (6,0) -- (6,3) -- (0,3) -- cycle;
 \draw[black,line width=2pt] (0,0) -- (3,0);
 \draw[black,line width=2pt] (0,3) -- (3,3);
 \draw[line width=3pt,red,densely dashed] (0,0.05) -- (6,0.05);
 \draw[line width=3pt,red,densely dashed] (0,2.95) -- (6,2.95);
  \node (ham1) at (10.5,1.5) {\scalebox{1}[-1]{\includegraphics[width=5.5cm]{ 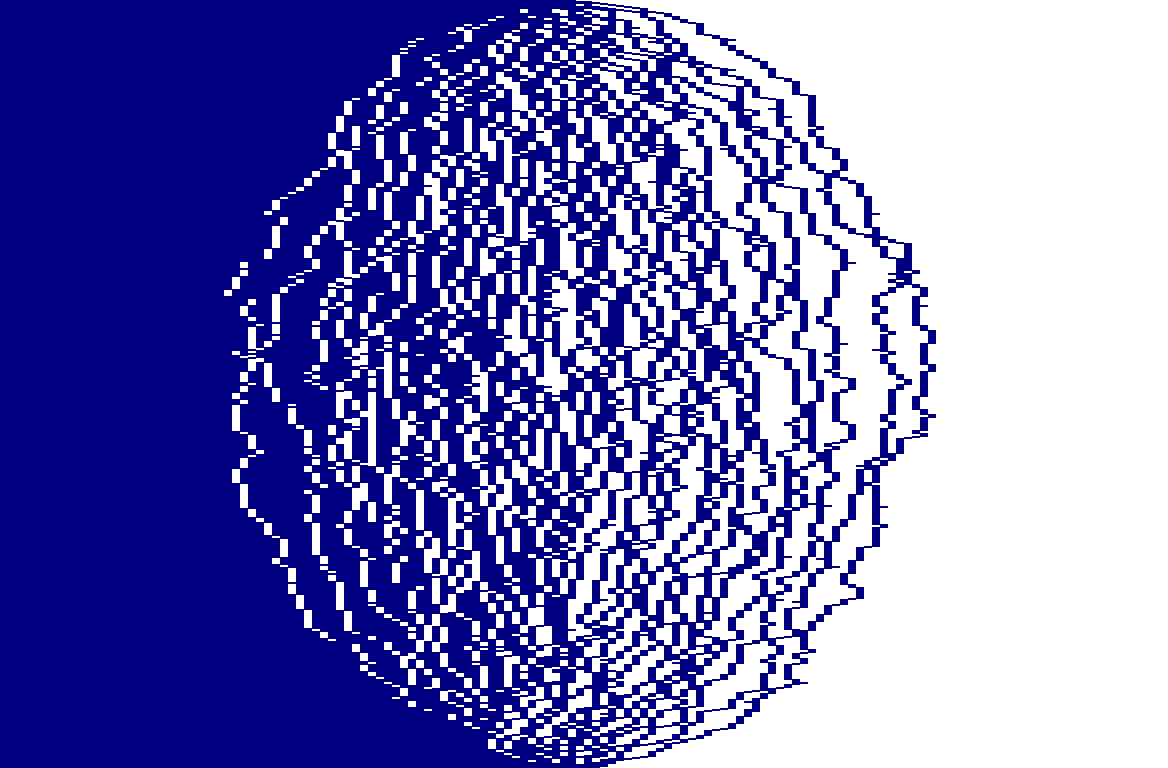}}};
 \end{scope}
\end{tikzpicture}
\caption{Six vertex model with domain wall boundary conditions in a square domain (shown in red dash). This model can be generated by considering a slab geometry and imposing the domain wall state at the top and bottom (shown in thick dark line). Taking the Hamiltonian limit amounts to sending $b\to0$ while scaling the vertical axis appropriately, as discussed in the text. From top to bottom, $b=1$, $b=1/2$, $b\to 0$. For a given $b$ we show the geometry on the left, a typical configuration for a somewhat large system at the free fermions point on the right. The degrees of Freedom are known to be frozen outside a disk in the scaling limit, by the arctic circle theorem \cite{jockusch1998random}. }
\label{fig:6v}
\end{figure}
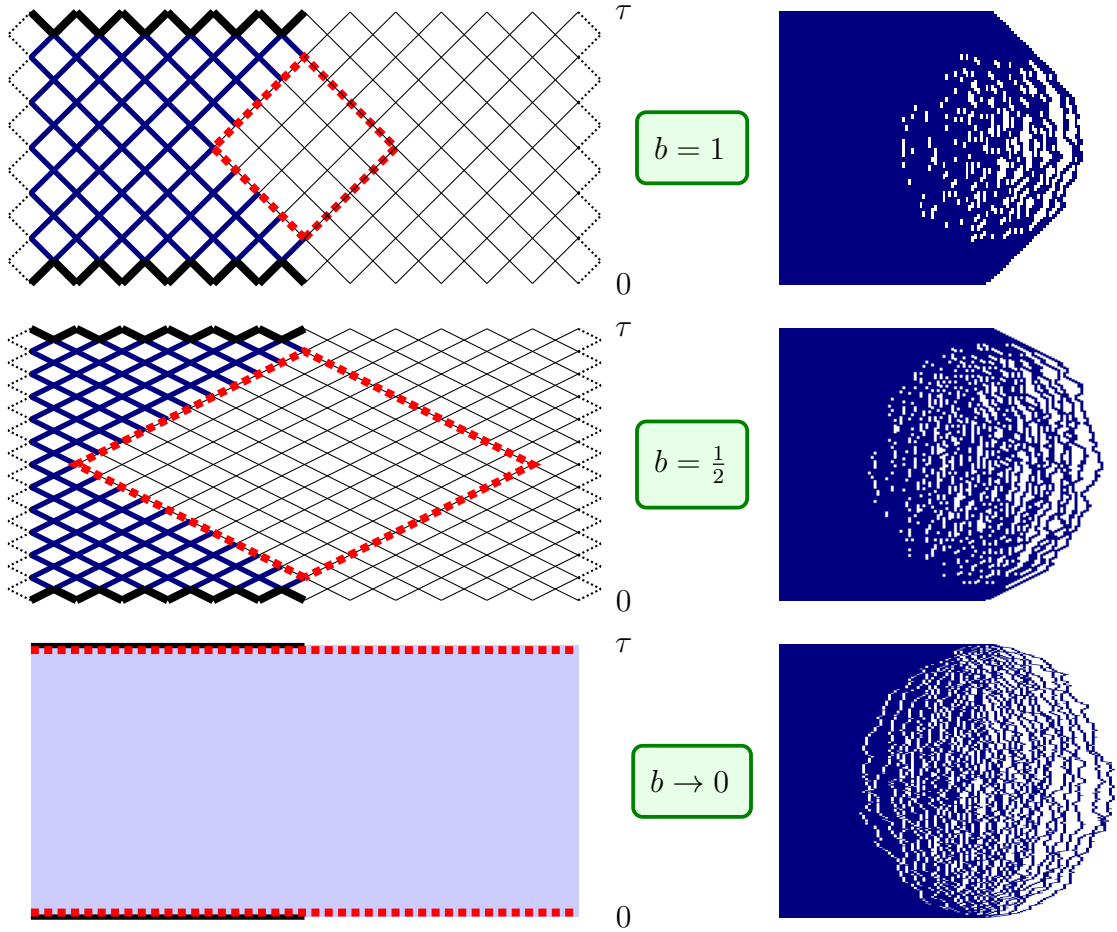

We choose to look at the model in a diagonal to diagonal direction, as shown in  figure~\ref{fig:6v}. 

We use the transfer matrix formalism, where imaginary time flows upwards in the picture. The transfer matrices corresponding this  problem are somewhat non-standard but nevertheless known, see e.g. \cite{Allegra_2016}. Define the operator
\begin{equation}\label{eq:Rmatrix}
    R_{j}=a\left[n_j n_{j+1}+h_j h_{j+1}\right]+b\left[\sigma_j^+\sigma_{j+1}^-+\sigma_{j+1}^+\sigma_j^-\right]+c\left[n_j h_{j+1}+h_j n_{j+1}\right].
\end{equation}
Here $n_j=\frac{1+\sigma_j^z}{2}$ is the particle number operator, while $h_j=1-n_j$ is the hole number operator.
$R_j$ implements the local rules for the six vertex model, and acts only on sites $j,j+1$: it is similar to the usual  $R-$matrix of the six vertex model, tensored with identities elsewhere. 
From this, one can build the two transfer matrices
\begin{align}
    T_{\rm e}=\prod_{j\,\textrm{even}} R_j\\
    T_{\rm o}=\prod_{j\,\textrm{odd}} R_j
\end{align}
which allow to compute any physical observable in the model. Now the observation of \cite{Allegra_2016} (see also \cite{Stephan_2014})
 is that imposing domain wall initial states at the top and bottom of the lattice reproduces a known setup, dubbed "six vertex model with domain wall boundary conditions (DWBC)" \footnote{Note that "domain wall" can have two different meanings. In the following, domain wall state will always refer to the state $\ket{\psi_0}$ which is imposed at the top and bottom, see figure \ref{fig:6v}. Domain wall boundary conditions will refer to the boundary conditions of the six vertex model shown inside the red dashed line in the same figure.}, for which several exact results are available. To be more precise, let us first set $a=1$. Then the partition function
 \begin{equation}
     Z_n(a=1,b,\Delta)=\braket{\psi_0|(T_{\rm e}T_{\rm o})^{2n}T_{\rm e}|\psi_0}
 \end{equation}
 coincides with the DWBC partition function, as can be seen in figure \ref{fig:6v} (top left). Now the relation with the spin chain (\ref{eq:H}) follows from the small $b$ expansion
 \begin{equation}
     R_j=1+b\left[\sigma_j^+\sigma_{j+1}^-+\sigma_j^-\sigma_{j+1}^++\frac{\Delta}{2}\left(\sigma_j^z\sigma_{j+1}^z-1\right)\right]+O(b^2),
 \end{equation}
 where the term proportional to $b$ is one element in the sum defining the Hamiltonian $H$ in (\ref{eq:H}). From the Trotter formula, it follows that
 \begin{equation}
     \lim_{b\to 0} \;(T_{\rm o}T_{\rm e})^{1/b}=e^{H},
 \end{equation}
 so 
 \begin{equation}\label{eq:hamlimit_Z}
     Z(\tau)=\lim_{n\to\infty} Z_n(a=1,b=\frac{\tau}{n},\Delta).
 \end{equation}

\subsection{Performing the Hamiltonian limit}
\label{sec:hamlimit}
\subsubsection{The partition function}
\label{sec:Z_limit}
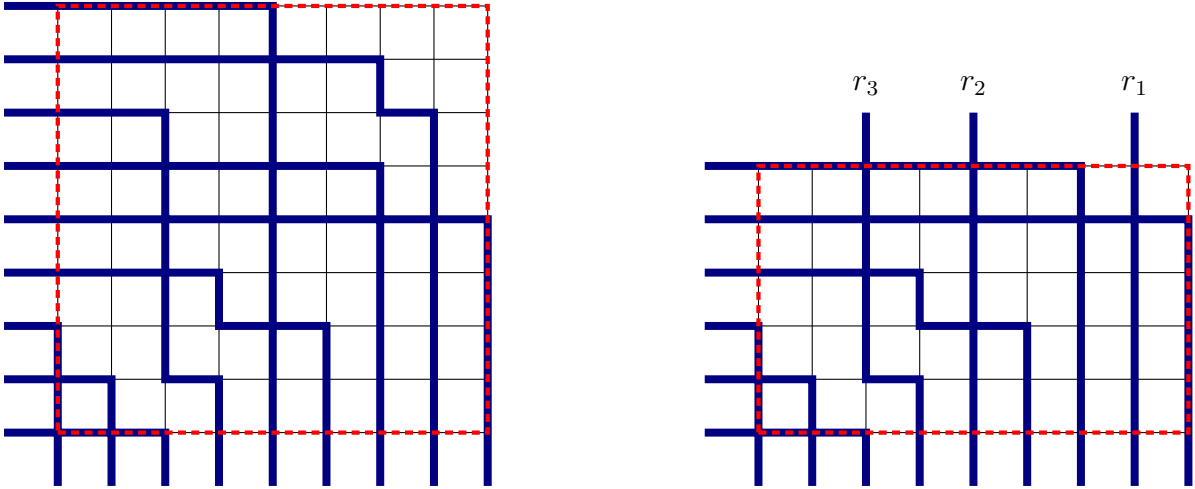
\begin{figure}[htbp]
\centering
\begin{tikzpicture}[xscale=-1,yscale=1,rotate=-45] 
\foreach \x in {0,1,2,3,4,5,6,7,8}{
\draw (5-0.5*\x,1+0.5*\x) -- (9-0.5*\x,5+0.5*\x);
\draw (5+0.5*\x,1+0.5*\x) -- (1+0.5*\x,5+0.5*\x);
}
 \draw[line width=3pt,color=mycolor] (5.5,0.5) -- (5,1) -- ++(-2,2) -- ++(0.5,0.5) -- ++(-1,1) -- ++(0.5,0.5) -- ++(-0.5,0.5) -- ++(1,1) -- ++ (-0.5,0.5) -- ++(2,2) -- ++(0.5,0.5);
	 \draw[line width=3pt,color=mycolor] (6,1) -- (5.5,1.5) -- ++(-2,2) -- ++(0.5,0.5) -- ++(-0.5,0.5) -- ++(1,1) -- ++(-1,1) -- ++(2,2) -- ++(0.5,0.5);
	 \draw[line width=3pt,color=mycolor] (6.5,1.5) --(6,2) -- ++(-2,2) -- ++(1,1) -- ++(-0.5,0.5) -- ++(1,1) -- ++(-0.5,0.5) -- ++(1,1) -- ++(0.5,0.5);
	 \draw[line width=3pt,color=mycolor] (7,2) -- (6.5,2.5) -- ++(-1,1) -- ++(0.5,0.5) -- ++(-1,1) -- ++(1,1) -- ++(-0.5,0.5) -- ++(1,1) -- ++(0.5,0.5);
	 \draw[line width=3pt,color=mycolor] (7.5,2.5) --(7,3) -- ++(-1,1) -- ++(0.5,0.5) -- ++(-0.5,0.5) -- ++(0.5,0.5) -- ++(-0.5,0.5) -- ++(1,1) -- ++(0.5,0.5);
	 \draw[line width=3pt,color=mycolor] (8,3) -- (7.5,3.5) -- ++(-0.5,0.5) -- ++(0.5,0.5) -- ++(-1,1) -- ++(1,1) -- ++(0.5,0.5);
	 \draw[line width=3pt,color=mycolor] (8.5,3.5) -- (8,4) -- ++(0.5,0.5) -- ++(-0.5,0.5) -- ++(0.5,0.5) -- ++(-0.5,0.5) -- ++(0.5,0.5);
	 \draw[line width=3pt,color=mycolor] (9,4) -- (8.5,4.5) -- ++(0.5,0.5) -- ++(-0.5,0.5) -- ++(0.5,0.5);
	 \draw[line width=3pt,color=mycolor] (9.5,4.5) -- (9,5) -- (9.5,5.5);
\draw[red,ultra thick,densely dashed] (9,5) -- (5,1) -- (1,5) -- (5,9) -- cycle;
\end{tikzpicture}\hfill
\begin{tikzpicture}[xscale=-1,yscale=1,rotate=-45] 
\draw (2.25,4.75) node {$r_1$};
\draw (3.75,6.25) node {$r_2$};
\draw (4.75,7.25) node {$r_3$};
\foreach \x in {0,1,2,3,4,5}{
\draw (5-0.5*\x,1+0.5*\x) -- (9-0.5*\x,5+0.5*\x);
}
\foreach \x in {0,1,2,3,4,5,6,7,8}{
\draw (5+0.5*\x,1+0.5*\x) -- (2.5+0.5*\x,3.5+0.5*\x);
}
 \draw[line width=3pt,color=mycolor] (5.5,0.5) -- (5,1) -- ++(-2,2) -- ++(0.5,0.5) -- ++(-1,1);
	 \draw[line width=3pt,color=mycolor] (6,1) -- (5.5,1.5) -- ++(-2,2) -- ++(0.5,0.5) -- ++(-0.5,0.5) -- ++(1,1) -- ++(-0.5,0.5);
	 \draw[line width=3pt,color=mycolor] (6.5,1.5) --(6,2) -- ++(-2,2) -- ++(1,1) -- ++(-0.5,0.5) -- ++(1,1) -- ++(-0.5,0.5);
	 \draw[line width=3pt,color=mycolor] (7,2) -- (6.5,2.5) -- ++(-1,1) -- ++(0.5,0.5) -- ++(-1,1) -- ++(1,1) -- ++(-0.5,0.5) -- ++(1,1) -- ++(0.5,0.5);
	 \draw[line width=3pt,color=mycolor] (7.5,2.5) --(7,3) -- ++(-1,1) -- ++(0.5,0.5) -- ++(-0.5,0.5) -- ++(0.5,0.5) -- ++(-0.5,0.5) -- ++(1,1) -- ++(0.5,0.5);
	 \draw[line width=3pt,color=mycolor] (8,3) -- (7.5,3.5) -- ++(-0.5,0.5) -- ++(0.5,0.5) -- ++(-1,1) -- ++(1,1) -- ++(0.5,0.5);
	 \draw[line width=3pt,color=mycolor] (8.5,3.5) -- (8,4) -- ++(0.5,0.5) -- ++(-0.5,0.5) -- ++(0.5,0.5) -- ++(-0.5,0.5) -- ++(0.5,0.5);
	 \draw[line width=3pt,color=mycolor] (9,4) -- (8.5,4.5) -- ++(0.5,0.5) -- ++(-0.5,0.5) -- ++(0.5,0.5);
	 \draw[line width=3pt,color=mycolor] (9.5,4.5) -- (9,5) -- (9.5,5.5);
\draw[red,ultra thick,densely dashed] (9,5) -- (5,1) -- (2.5,3.5) -- (6.5,7.5) -- cycle;
\end{tikzpicture}
\caption{\emph{Left:} Six vertex with domain wall boundary condition, that is paths entering a square (dashed red) from the bottom and leaving from the left. This model is exactly the one shown in figure \ref{fig:6v}. In the Hamiltonian limit, the partition function becomes exactly the partition function $Z(\tau)$ in the spin chain. \emph{Right:} Same model with a finite number $l$ of rows removed, but  $l$ paths are allowed to exit from the top at fixed locations $r_1,\ldots,r_l$ ($l=3$ in the picture). In the Hamiltonian limit, the ratio of the two partition functions converges to the amplitudes we are after, in the XXZ spin chain.}
\label{fig:6v_dwbc}
\end{figure}

Let us first discuss the partition function, which illustrates the main difficulties in taking the series of limits necessary for our purposes. The partition function $Z_n(a,b,\Delta)$ was considered in \cite{Korepin1982} and computed as a determinant in \cite{Izergin1987,IzerginCokerKorepin1992}. This was done by  considering an inhomogeneous generalization, with weights at the intersection of the $k$-th row and $l-$th column (see figure \ref{fig:6v_dwbc}) given by 
\begin{equation}\label{eq:inh_weights}
    a_{kl}= \sin(\lambda_k+\mu_l)\quad,\quad b_{kl}=\sin(\lambda_k-\mu_l) \quad,\quad c_{kl}=\sin\gamma \quad,
\end{equation}
with $\{\lambda_k\}_{1\leq k\leq N}$ and $\{\mu_k\}_{1\leq k\leq N}$  two sets of real numbers (we assume $|\Delta|<1$ for now).

This generalization makes it more convenient to apply the Quantum Inverse Scattering Method \cite{korepin_bogoliubov_izergin_1993}. Indeed, one can show that the partition function satisfies certain recursion relations stemming from the Yang-Baxter algebra which specify it uniquely\cite{Korepin1982}. It was shown later \cite{Izergin1987,IzerginCokerKorepin1992} that the following partition function
\begin{align}\label{eq:IK_formula}
    Z_N(\{\lambda_k\},\{\nu_l\})=\frac{\prod_{k,l=1}^N \sin(\lambda_k-\nu_l+\gamma)\sin(\lambda_k-\nu_l)
    }{\prod_{k<k'}\sin(\lambda_k-\lambda_{k'})\prod_{l<l'}\sin(\nu_l-\nu_{l'})}
    \;\det_{1\leq k,l\leq N} \left(\phi(\lambda_k-\nu_l)\right)
\end{align}
satisfies the aforementioned recursion relations, so is the partition function we are after. Here $\phi$ is the function defined by
\begin{equation}\label{eq:weight_fct}
    \phi(\epsilon)=
    \int_{\mathbb{R}}w_\epsilon(x)\,dx\qquad,\qquad w_\epsilon(x)=e^{-\epsilon x} \frac{1-e^{-\gamma x}}{1-e^{-\pi x}}.
\end{equation}
One can compute $\phi(\epsilon)$, but the integral form is more convenient for our purposes. 
The next step is to take the homogeneous limit $\lambda_k\to\lambda$, $\nu_k\to\nu$. In this limit the prefactor vanishes while the determinant blows up. As explained in \cite{IzerginCokerKorepin1992}, it can nevertheless be performed by appropriate row column manipulations, and using l'Hospital's rule repeatedly. Rescaling all vertex weights by a global factor $d$ amounts to multiplying the partition function by a factor $d^{N^2}$, so we may use that to set $a=1$ as needed.
The final result reads
\begin{equation}
    Z_N=\frac{\left[\sin\epsilon\right]^{N^2}}{\prod_{k=0}^{N-1}k!^2} \det_{0\leq j,k\leq N-1} \left(\frac{d^{i+j}\phi(\epsilon)}{d\epsilon^{i+j}}\right),
\end{equation}
with the choice $\lambda=\epsilon+\gamma/2$, $\mu=\gamma/2$, $d=\frac{1}{\sin(\gamma+\epsilon)}$, that is
\begin{equation}
    a=1\qquad,\qquad b=\frac{\sin\epsilon}{\sin(\gamma+\epsilon)}\qquad,\qquad \Delta=\cos\gamma
\end{equation}
with $\epsilon>0$, $0<\gamma<\pi-\epsilon$. 
Accessing $Z(\tau)$ amounts to taking a double scaling limit involving $\epsilon\to 0$, see (\ref{eq:hamlimit_Z}). It is more complicated, but can be done by rewriting \cite{Slavnov}
\begin{equation}\label{eq:Z_rewrite}
    Z_N=\left[\frac{\sin\epsilon}{\epsilon}\right]^{N^2} \frac{\underset{0\leq j,k\leq N-1}{\det} \left(\int_{\mathbb{R}} x^{i+j}w_\epsilon(x)dx\right)}{\underset{0\leq j,k\leq N-1}{\det} \left(\int_{\mathbb{R}} x^{i+j} e^{-\epsilon x} dx\right)}.
\end{equation}
The matrix in the denominator is sufficiently simple that its determinant can be evaluated (explaining in passing why (\ref{eq:Z_rewrite}) holds), but its inverse is also known. Using $\det A/\det B=\det(B^{-1}A)$ then allows to rewrite the partition function is such a way that the Hamiltonian limit $b=\tau/N$ and $N\to\infty$ can be performed. We quote only the final result \cite{Stephan_return} below
\begin{equation}\label{eq:Z_theexact}
    Z(\tau)=\exp\left(-\frac{\tau^2\sin^2\gamma}{6}\right)\det_{L^2(\mathbb{R})}(I-V_{\tau\sin\gamma})
\end{equation}
where appears on the rhs the Fredholm determinant of an operator with kernel
\begin{equation}\label{eq:kern_main}
     V_{\tau}(x,y)=\frac{\sqrt{y}J_0(2\sqrt{x})J_0'(2\sqrt{y})-\sqrt{x}J_0(2\sqrt{y})J_0'(2\sqrt{x})}{(x-y)} \left[\Theta(y)-w_0\left(\frac{y}{\tau}\right)\right]
\end{equation}
acting on functions in $L^2(\mathbb{R})$ (see (\ref{eq:fred_def}) for a definition).  Here $\Theta$ denotes the Heaviside step function, $\Theta(x)=1$ for $x>0$ and $\Theta(x)=0$ for $x\leq 0$.

Let us now summarize how this result was obtained. The inhomogeneous partition function (\ref{eq:IK_formula}) follows from the Yang-Baxter  integrability of the six vertex model. Two successive limits are then necessary to access $Z(\tau)$. The first is a homogeneous limit. In this limit one obtains the determinant of a matrix whose elements depend only on the sum $i+j$ of the row and column indices. Such Hankel matrices are well-known to be related to the theory of orthogonal polynomials\cite{Szego}. Taking the final Hamiltonian (or Trotter) limit requires working with orthogonal polynomials; the full derivation can be found in  the self-contained \ref{app:orthopolys}.
\subsubsection{The one-particle amplitude}
\label{sec:amplitude_limit}
The one-particle amplitude $A_x(\tau)$ and its generating function $h(\tau|z)$ can be obtained using the exact same strategy as explained above. On the six vertex side, the analogous one-point boundary was studied in Ref~\cite{Bogoliubov_2002}. Observe that in the first row (starting from the top in figure \ref{fig:6v_dwbc}), there will be exactly one vertical lattice path. Denote by $H_N^{(r)}(\epsilon)$ the probability that this vertical path  lies at site $r$, counting from the right. For later convenience we write explicitly the dependence on $\epsilon$, but not $\gamma$. Make a generating function out of it:
\begin{equation}
    h_N(\epsilon|w)=\sum_{r=1}^N H_N^{(r)}(\epsilon)w^{r-1}.
\end{equation}
The first observation is that the desired function $h(\tau|z)$ is given by the following limit
\begin{equation}\label{eq:limit_h}
    h(\tau|z)=\lim_{N\to\infty} h_N\left(\frac{\tau\sin\gamma}{N}\right|\left.\frac{zN}{\tau}\right).
\end{equation}
Alternatively, this means $\left[\frac{N}{\tau}\right]^{r-1} H_N^{(r)}(\frac{\tau\sin\gamma}{N})\to A_{r-1}(\tau)$ as $N\to\infty$. 
This can be shown by expressing $H_N^{(r)}(\epsilon)$ using the diagonal to diagonal transfer matrix discussed in section  \ref{sec:relation}, and performing the Hamiltonian limit. The relation  is useful because an exact determinant expression is available for $h_N(\epsilon|w)$. Introduce the moments
\begin{equation}
    \label{eq:moments}
    \braket{x^k}_\epsilon=\int_{\mathbb{R}} x^k w_\epsilon(x)\,dx,
\end{equation}
where $w_\epsilon$ is given by (\ref{eq:weight_fct}). With this notation, the determinant entering the partition function (\ref{eq:Z_rewrite}) reads, up to a sign,
\begin{equation}
    D_N=\det_{0\leq i,j\leq N-1}\left(\braket{ x^{i+j}}_\epsilon\right).
\end{equation}
Let us now quote the exact result of Ref.\cite{Bogoliubov_2002,ColomoPronkocurve} for $h_N$ already in the homogeneous limit, while referring to those works for a detailed derivation starting from the inhomogeneous generalization:
\begin{equation}
 h_N(\epsilon|\Gamma(s))=\left[\frac{\sin s \sin(s+\gamma+\epsilon)}{\sin(\gamma+\epsilon)}\left(\frac{\sin(s+\epsilon)}{\sin s }\right)^N\right]\times \left[\frac{(N-1)!}{\sin^N \epsilon}\frac{\tilde{D}_N(s)}{D_N}\right],
\end{equation}
where
\begin{equation}\label{eq:Gamma}
\Gamma(s)=\frac{\sin(\gamma+\epsilon)\sin(s+\epsilon)}{\sin\epsilon\sin(\gamma+s+\epsilon)},
\end{equation}
and 
\begin{equation}
    \tilde{D}_n(s)=\left|
    \begin{array}{ccccc}
    \braket{x^0}_\epsilon&\braket{x}_\epsilon&\ldots&\braket{x^{n-1}}_\epsilon\\
    \braket{x}_\epsilon&\braket{x^2}_\epsilon&\ldots&\braket{x^{n}}_\epsilon\\
    \vdots &&&\vdots\\
    \braket{x^{n-2}}_\epsilon&\braket{x^{n-1}}_\epsilon&\ldots&\braket{x^{2n-1}}_\epsilon\\
    \braket{x^0}_{\epsilon+s}&\braket{x}_{\epsilon+s}&\ldots&\braket{x^{n-1}}_{\epsilon+s}
    \end{array}
    \right|.
\end{equation}
The last determinant is related to the polynomial orthogonal with respect to the weight function $w_\epsilon$ as is explained in \ref{app:ortho_elementary}. This means one may rewrite the exact generating function as follows
\begin{equation}\label{eq:hn_orthorewrite}
    h_N(\epsilon|\Gamma(s))=\frac{\sin s \sin(s+\gamma+\epsilon)\sin^N(s+\epsilon)}{\sin(\gamma+\epsilon)\sin^N s \sin^N \epsilon}(N-1)!\sqrt{\frac{\epsilon D_{N-1}}{D_N}}\int_{\mathbb{R}}\frac{p_{N-1}(x)}{\sqrt{\epsilon}}w_{\epsilon+s}(x)\,dx
\end{equation}
where the $p_k$ are polynomials of degree $k$ which are orthonormal with respect to the weight function $w_{\epsilon}$, $\braket{p_k|p_q}_\epsilon=\delta_{kq}$. We write again  $p_k(x)=p(k,\epsilon|x)$ in the following, to emphasize the implicit dependence on $\epsilon$, which is key for our purpose. Now the Hamiltonian limit is obtained by setting $\epsilon=\frac{\tau\sin\gamma}{N}$, and sending $N$ to infinity, see (\ref{eq:limit_h}). Most terms are easy, except for the orthonormal polynomials. Fortunately, it can be shown that the limit
\begin{equation}
    q(\alpha|x)=\lim_{N\to\infty} \sqrt{\frac{N}{\alpha}}p(N,\alpha/N|x)
\end{equation}
is well defined, satisfies an exact differential equation with a source term $f(\tau)$ determined by a Fredholm determinant, see equations (\ref{eq:ortho_ode},\ref{eq:ortho_f},\ref{eq:ortho_det},\ref{eq:ortho_kernel}). The whole \ref{app:orthopolys} is devoted to a self-contained  derivation of this result. At least the way we are proceeding, the main technical difficulties lie in establishing the aforementioned ODE for $q(\alpha|x)$. With this result at hand however, the Hamiltonian limit may be performed without any other difficulties. We obtain
\begin{equation}
    \label{eq:halphabeta}
 h\left(\frac{\tau}{\sin\gamma}\left|\frac{\sin s}{\sin(\gamma+s)}\right)\right.=\frac{\sin s \sin(\gamma+s)}{\sin\gamma} e^{\alpha \cot s}\times
 g(\tau|s),
\end{equation}
where
\begin{equation}
    g(\tau|s)=\int_{\mathbb{R}}q(\alpha|x)w_{s}(x)\,dx.
\end{equation}
Using the ODE (\ref{eq:ortho_ode}) for $q(\alpha|x)$, one can check that $g$ satisfies the PDE
\begin{equation}\label{eq:exact_pde_bis}
    \left[\tau\partial_\tau^2+\partial_\tau+f(\tau)\right]g(\tau|s)=\partial_s g(\tau|s),
\end{equation}
where $f(\tau)$ is given by (\ref{eq:ortho_f}). 
The PDE (\ref{eq:exact_pde_bis}) translates into a PDE for $h$ using  (\ref{eq:halphabeta}). Coming back to the original variable $z=\frac{\sin s}{\sin(\gamma+s)}$ and using the formula for the exact partition function finally yields the result (\ref{eq:pde}),(\ref{eq:Y}) claimed in the introduction. Even though the present derivation holds only for vertex weights parametrized by trigonometric functions (implying $|\Delta|<1$), it can be adapted to handle other regimes as well, and our main result (\ref{eq:pde},\ref{eq:Y}) --which is written without reference to a specific parametrization--  unsurprisingly also holds for arbitrary values of $\Delta$.

\subsubsection{The multi-particle amplitudes}
\label{sec:amplitudes_limit}
The strategy of the previous subsection can be generalized to tackle the multiparticle case. On the six vertex side, there are exactly $l$ vertical paths in the $l-$th row, starting again from the top. Denote by $Z_N^{(r_1,\ldots,r_s)}$ the partition function on a $N\times(N-l)$ lattice with domain wall boundary conditions, but with $l$ extra $c$-vertices inserted at the top, as is shown in figure \ref{fig:6v_dwbc}.

There is a multiple contour integral for this partition function, which is given by
\begin{equation}\label{eq:cp_Zbottom}
    \frac{Z_N^{r_1,\ldots,r_l}}{Z_N}=\frac{1}{a^{l(N-1)}c^l} \oint_{\mathcal{C}^l} \prod_{j=1}^l \frac{dz_j}{2\ci\pi z_j^{r_j}}\prod_{1\leq j<k\leq l} \frac{z_k-z_j}{z_j z_k-2\Delta z_j+1} h_N\left(\epsilon\left|\frac{a z_1}{b},\ldots ,\frac{a z_l}{b}\right)\right. ,
\end{equation}
where
\begin{equation}
 h_{N}(\epsilon|z_1,\ldots,z_l)=\frac{1}{\prod_{1\leq j<k\leq l}(z_k-z_j)}\det_{1\leq j,k\leq l}\left(z_k^{l-j}(z_k-1)^{j-1}h_{N-l+j}(\epsilon|z_k)\right).
\end{equation}
The formula holds in the homogeneous case. We refer to Refs.\cite{ColomoPronkotopbottom,ColomoDiGiulioPronko} for a detailed derivation, and discussions of the many subtleties involved in establishing such a formula from the inhomogeneous model. 
Setting $a=1$, $b=\tau/N$ and taking the limit $N\to\infty$, the lhs of (\ref{eq:cp_Zbottom}) yields exactly the amplitude $A_{r_1,\ldots,r_l}(\tau)$. 
Hence the only calculation left is that of the limit
\begin{equation}\label{eq:multih_limit}
    h(\tau|z_1,\ldots,z_l)=\lim_{N\to\infty} h_{N}\left.\left(\frac{\tau\sin\gamma}{N}\right|\frac{z_1 N}{\tau},\ldots,\frac{z_l N}{\tau} \right),
\end{equation}
which can be done as follows. 
Write
\begin{equation}
    h_{N}\left.\left(\frac{\tau\sin\gamma}{N}\right|\frac{z_1 N}{\tau},\ldots,\frac{z_l N}{\tau} \right)
    =\frac{1}{\prod_{j<k} (z_k-z_j)}\det_{1\leq j,k\leq l}\left(\mathcal{H}_{N,j}(\tau|z_k)\right)
\end{equation}
where the matrix elements are determined by
\begin{equation}
    \mathcal{H}_{N,j}(\tau|z)=z^{m-j}\left(\frac{Nz}{\tau}-1\right)^{j-1}h_{N-m+j}\left.\left(\frac{\tau\sin\gamma}{N}\right|\frac{Nz}{\tau}\right).
\end{equation}
The matrix elements blow up when $N\to\infty$, unless $j=1$. However the determinant stays finite, because the diverging part in a given row becomes proportional to other rows in the limit, resulting in cancellations. The first line is simple:
\begin{equation}
    \lim_{N\to\infty} \mathcal{H}_{N,1}(\tau|z)=z^{l-1}h(\tau|z).
\end{equation}
The second line can be regularized by subtracting $N/\tau$ times the first line without affecting the determinant:
\begin{align}\nonumber
    \mathcal{H}_{N,2}-\frac{N}{\tau}\mathcal{H}_{N,1}&=\frac{Nz^{l-1}}{\tau}\left[h_{N-l+2}\left.\left(\frac{\tau\sin \gamma}{N}\right|\frac{zN}{\tau}\right)-h_{N-l+1}\left.\left(\frac{\tau\sin \gamma}{N}\right|\frac{zN}{\tau}\right)\right]-z^{l-2}h_{N-l+2}\\
    &\sim z^{l-1}\partial_\tau h(\tau|z)-z^{l-2}h(\tau|z)
\end{align}
as $N\to\infty$. To obtain the last formula, we have made the  assumption that the form $h_{N}(\epsilon|z)=\sum_{k=0}^{N-1}(\epsilon N)^k u_k(z)$ holds for large $N$ and small $\epsilon$, where $u_k$ is a polynomial in $z$. Similarly, we obtain
\begin{equation}
    \sum_{p=0}^{j-1} C_j^p\left(-\frac{N}{\tau}\right)^{p}\mathcal{H}_{N,j-p}\sim z^{l-j}[z\partial_\tau-1]^{j-1} h(\tau|z)
\end{equation}
which leads to 
\begin{equation}
    \label{eq:final_multih}
    h(\tau|z_1,\ldots,z_l)=\frac{1}{\prod_{1\leq j<k\leq l}(z_j-z_k)} \det_{1\leq j,k\leq l}\left(z_k^{l-j}[1-z_k\partial_\tau]^{j-1}h(\tau|z)\right).
\end{equation}
Setting $a=1$, $b=\tau/N$, and inserting this limit in (\ref{eq:cp_Zbottom}) yields
\begin{equation}
    \label{eq:multiamplitude_bis}
    A_{x_l,\ldots,x_1}(\tau)=\oint_{\mathcal{C}^l} \prod_{j=1}^l\frac{dz_j}{2\ci\pi z_j^{x_j+l}} \prod_{1\leq j<k\leq l}\frac{z_j-z_k}{z_j z_k-2\Delta z_k+1} h(\tau|z_1,\ldots,z_l)
\end{equation}
which is identical to the contour integral (\ref{eq:multiamplitude}) written in the introduction. 

Let us finally note that at various stages in the derivation, we have assumed certain scaling forms for $h_N(\epsilon|z)$ and related objects. These scaling forms are natural but would in principle require stronger justifications. To alleviate this issue, several very strong numerical checks are presented in \ref{app:checks}.

\section{Examples}
\label{sec:examples}
We discuss in this section various simple examples where our main formulae turn out to be relevant. Consider the Hamiltonian
\begin{equation}\label{eq:newH}
    H_{XXZ}=\sum_{j\in\mathbb{Z}}\left(S_{j}^x S_{j+1}^x+S_{j}^y S_{j+1}^y+\Delta  \left[S_j^z S_{j+1}^z-\frac{1}{4}\right]\right)
\end{equation}
with spin operators $S_j^\alpha=\frac{1}{2}\sigma_j^\alpha$. This Hamiltonian is nothing but the one in (\ref{eq:H}) up to a factor $2$, $2H_{XXZ}=H$. We use this new Hamiltonian (\ref{eq:newH}) as it is more convenient to make contact with existing literature, e.g. \cite{antal1999transport,ColluraDeLucaViti,Allegra_2016,Stephan_return}. In particular, the maximum speed of propagation is $v_{\rm max}=1$ with the normalization (\ref{eq:newH}). 

The first class of problems of interest includes imaginary-time setups for a local operator $\hat{O}$:
\begin{equation}\label{eq:imaginarytimeevolution}
    \braket{\hat{O}}_{R,y}=\frac{\braket{\psi_0|e^{(R-y)H_{XXZ}} \hat{O} e^{(R+y)H_{XXZ}}|\psi_0}}{\braket{\psi_0|e^{2RH_{XXZ}}|\psi_0}},
\end{equation}
where $R>0$, and $y\in[-R,R]$. This can be interpreted as a statistical model in a slab geometry, where the vertical direction is continuous ($y\in[-R,R]$) but the horizontal one is discrete ($x\in\mathbb{Z}$). The second class is that of a quantum quench:
\begin{equation}
    \braket{\psi(t)|\hat{O}|\psi(t)}=\braket{\psi_0|e^{\ci t H_{XXZ} t}\hat{O}e^{-\ci t H_{XXZ} }|\psi_0}
\end{equation}
which has been widely studied in recent years \cite{antal1999transport,antal_2008,ColluraDeLucaViti,YoungItalians,Eisler}. In this case $\ket{\psi(t)}$ solves the Schr\"odinger equation with initial state $\ket{\psi_0}$.  We have less to say about this topic in this section, but will nevertheless comment on it at various moments. 

A simple strategy to exploit (\ref{eq:pde},\ref{eq:multiamplitude}) is to use the remark already made in the introduction that everything is ultimately determined by the partition function. Except at the free fermions point, the Fredholm determinant defining the partition function cannot be computed explicitly: since it enters (\ref{eq:pde}) as a source term, solving the PDE exactly is a hopeless  task.  However, the asymptotics (as $|\tau|\to\infty$) of the partition function are available in most regimes, which means one can access $h(\tau|z)$ for large $|\tau|$, and exploit this to obtain the large $|\tau|$ behavior of more complicated observables. 

For example, if $h(\tau|z)$ takes the form (we take $\simeq$ to mean either exact or asymptotic equality, without being precise at the moment)
\begin{equation}\label{eq:h_ideal}
    h(\tau|z)\simeq e^{\tau F(z)+G(z)},
\end{equation}
then the determinant entering (\ref{eq:multiamplitude}) reduces to a Vandermonde determinant, which is easy to compute. In this case, the amplitude simplifies to
\begin{equation}
    \label{eq:multiamplitudes_ideal}
    A_{x_l,\ldots,x_1}(\tau)\simeq \oint_{\mathcal{C}^l} \prod_{j=1}^l \frac{dz_j h(\tau|z_j)}{2\ci\pi z_j^{x_j+l}}\,
    \prod_{1\leq j<k\leq l} \frac{z_j-z_k+z_j z_k(F(z_j)-F(z_k))}{z_j z_k-2\Delta z_k+1}.
\end{equation}
This identity will be exploited in subsections \ref{sec:gapped}, \ref{sec:freefermions}, \ref{sec:trigo}.

\subsection{The gapped case $\Delta>1$}
\label{sec:gapped}
We start with the gapped regime $\Delta>1$ in imaginary time, which is arguably simplest. We use the parametrization $\Delta=\cosh \eta$ for some $\eta>0$. The most suitable form of the exact partition function in this regime reads \cite{Stephan_return} (see also \cite{ColomoPronkodet})
\begin{equation}\label{eq:fred_discrete}
    Z(\tau)=e^{\tau\sinh\eta}\det_{0\leq j,l\leq \infty}(\delta_{jl}-V_{jl})
\end{equation}
with a kernel determined from a confluent hypergeometric function: 
\begin{equation}\label{eq:kern_discrete}
    V_{jl}=e^{-(j+l)\eta}\int_{0}^{2\tau\sinh \eta}\, _{1}F_1(-j,1,\alpha)\, _{1}F_1(-l,1,\alpha) e^{-\alpha}d\alpha.
\end{equation}
The Fredholm determinant is that of an operator which acts on sequences in $\ell^2(\mathbb{N})$. 
One can check that $V_{00}=1-e^{-2\tau \sinh \eta}$, $V_{jj}\sim e^{-2\eta j}$ up to exponentially small corrections as $\tau\to\infty$, and that off-diagonal elements are also exponentially small. This means
\begin{equation}
    Z(\tau)=e^{-\tau\sinh \eta} \left[\prod_{k=0}^\infty \left(1-e^{-2k\eta}\right)+\varepsilon(\tau)\right]
\end{equation}
where $\varepsilon(\tau)$ is exponentially small, so can be safely ignored. This is what we do in the following. We obtain $Q(\tau)=-\sinh\eta$ and the PDE reduces to
\begin{equation}\label{eq:pde_gapped}
    \left(\tau\partial_\tau^2+\left[1-2\tau\left(\frac{1}{z}-\cosh\eta\right)\right]\partial_\tau+e^{-\eta}-z\right) h(\tau|z)=\left(1-2z\cosh\eta+z^2\right)\partial_z h(\tau|z).
\end{equation}
Making the Ansatz $h(\tau|z)=e^{\tau F(z)+G(z)}$, there is a trivial solution with $F(z)=0$, which implies
\begin{equation}
    G'(z)=\frac{1}{e^{\eta}-z}.
\end{equation}
Exponentiating and fixing the initial condition $h(\tau|0)=1$ yields
\begin{equation}\label{eq:hab_gapped}
    h(\tau|z)=\frac{1}{1-ze^{-\eta}}.
\end{equation}
The fact that the generating function does not scale with $\tau$ is expected in this gapped regime, where the highly attractive interactions prevent the particles from moving far to the right, and imaginary time evolution simply projects to the ground state of the gapped XXZ spin chain with kink boundary conditions  \cite{Nachtergaele2007,Alcaraz_kink,Dijkgraaf_2009,reshetikhin_domainwallxxz}. Now (\ref{eq:hab_gapped}) is obviously of the form (\ref{eq:h_ideal}), which means the amplitudes read
\begin{equation}
    A_{x_1,\ldots,x_l}(\tau)=\oint_{\mathcal{C}^l} \prod_{j=1}^l \frac{dz_j}{2\ci \pi z_j^{x_j+l}(1-e^{-\eta} z_j)}\prod_{1\leq j<k\leq l}\frac{z_j-z_k}{z_j z_k-2\Delta z_k+1}
\end{equation}
for large $\tau$, with exponentially small corrections. Carefully evaluating the residues on the rhs, we obtain
\begin{equation}
     A_{x_l,\ldots,x_1}(\tau)=e^{-\eta \sum_{j=1}^l\left( x_j+j-1\right)}
\end{equation}
for large $\tau$, where recall $-l+1\leq x_l<\ldots<x_1$. The probability of observing the configuration with particles at $x_l,\ldots,x_1$ at imaginary time $y$ (see (\ref{eq:imaginarytimeevolution})) is
\begin{equation}
    P_{x_l,\ldots,x_1}(y,R)=A_{x_l,\ldots,x_1}\left(\frac{R-y}{2}\right) A_{x_l,\ldots,x_1}\left(\frac{R+y}{2}\right)
    \frac{Z\left(\frac{R-y}{2}\right)Z\left(\frac{R+y}{2}\right)}{Z(R)}.
\end{equation}
Using the asymptotic formula for the amplitudes, we obtain
\begin{equation}\label{eq:probas_gapped}
    P_{x_l,\ldots,x_1}(y,R)=\prod_{n=1}^\infty \left(1-q^n\right)q^{\sum_{j=1}^l\left( x_j+j-1\right)},
\end{equation}
where 
\begin{equation}\label{eq:q}
q=e^{-2\eta}=(\Delta-\sqrt{\Delta^2-1})^2.
\end{equation}
Equation (\ref{eq:probas_gapped}) holds when both $R-y$ and  $R+y$ are large. 
 The fact that the probabilities are normalized corresponds to the statement that $q\mapsto \prod_{n=1}^{\infty}\frac{1}{1-q^n}$ is the generating function of integer partitions, and the known bijection between partitions and fermionic configurations obtained from $\ket{\psi_0}$ by a finite number of particle moves.

Let us now give two examples of physical observables. The probability that there are $n$ particles in $\mathbb{Z}_{\geq k}$ is
\begin{align}
    \mathcal{P}_{n,k}&=\sum_{k\leq x_n<\ldots<x_1}
    \left(P_{x_n,\ldots,x_1}+\sum_{m=1}^\infty 
    \sum_{-n+m+1<x_{n+m}<\ldots<x_{n+1}<k} P_{x_{n+m},\ldots,x_1}\right)\\
    &=q^{n(k+n-1)} \frac{(q,q)_\infty}{(q,q)_n (q,q)_{n+k-1}}\label{eq:probas_q}
\end{align}
where we have used the $q-$Pochhammer symbol notation $(a,q)_n=\prod_{k=0}^{n-1}(1-a q^k)$. The probability that there are no particles, also known as emptiness formation probability (EFP) is  $\mathcal{P}_{0,k}=\prod_{j=k}^\infty(1-q^j)$. The mean particle number in $\mathbb{Z}_{\geq k}$ is $N_{\geq k}=\sum_{n\geq 0} \mathcal{P}_{n,k}$, and the density at site $k$ reads \begin{equation}
    \label{eq:density_gapped}
    \rho_k=\sum_{n\geq 0} n(\mathcal{P}_{n,k}-\mathcal{P}_{n,k+1}).
\end{equation}
This density profile decreases from one to zero as $k$ is increased, smoothening the initial domain wall density. The limit $q\to 1^{-}$ corresponds to a famous problem in quantum statistical mechanics. Introducing the scaled variable $u=(k-1/2)\log(1/q)$, the density profile can be shown to converge to the celebrated Fermi-Dirac distribution $u\mapsto \frac{1}{e^{u}+1}$.

Let us now comment on the analogous real time problem, $\tau=\ci t$, $t \geq 0$, which also does not support transport on macroscopic scales \cite{Gobert} due to energy considerations \cite{Misguich_numerics}. The return probability $R(t)=|Z(\ci t)|^2$ has been first investigated in \cite{MosselCaux}, which showed that its time average satisfies $\overline{R(t)}\geq \prod_{n\geq 1}(1-q)^2$, and presented numerical evidence that the bound is tight. As noted in \cite{Stephan_return}, the formula looks plausible from the exact Fredholm determinant (\ref{eq:fred_discrete}),(\ref{eq:kern_discrete}), simply by arguing that the diagonal term dominate. However, it is not clear from the formula whether oscillations --on top of the mean value-- persist, or slowly decay at extremely large times as argued in \cite{MosselCaux}. If so, then equations (\ref{eq:density_gapped}),(\ref{eq:probas_q}),(\ref{eq:q}) provide an exact formula for the density profile at extremely large times, simply by performing the exact same calculations as above. If not, then there are constructive interference effects, and one probably just gets a lower bound. 

\subsection{The free fermions case $\Delta=0$}
\label{sec:freefermions}
Our second example is the only one for which the partition function is simple and explicit even for finite $\tau$. It is given by (see e.g. \cite{BorodinOkounkov,prahofer2002scale}, or taking the Hamiltonian limit of the known partition function $Z_n(a,b,\Delta=0)=(1+b^2)^{N^2/2}$ in the six vertex model): 
\begin{equation}
    Z(\tau)=e^{\tau^2/2}.
\end{equation}
This implies the PDE
\begin{equation}
    \left(\tau\partial_\tau^2+\left[1-\frac{2\tau}{z}\right]\partial_\tau+3\tau-z\right)h(\tau|z)=(1+z^2)\partial_z h(\tau|z),
\end{equation}
which has a very simple exact solution
\begin{equation}\label{eq:h_free}
    h(\tau|z)=e^{\tau z}.
\end{equation}
The result is once again of the form (\ref{eq:h_ideal}), which implies the contour integral formula
\begin{align}\label{eq:P_rmt}
A_{x_l,\ldots,x_1}(\tau)&= \oint_{\mathcal{C}^l} \prod_{j=1}^l \frac{dz_j e^{\tau z_j}}{2\ci\pi z_j^{x_j+l}}\prod_{1\leq j<k\leq l}(z_j-z_k)\\\label{eq:P_det}
&=\det_{1\leq j,k\leq l}\left(\oint_{\mathcal{C}} \frac{dz\,e^{\tau z}}{2\ci\pi z^{x_j+k}}\right)\\
&=\tau^{\sum_{j=1}^l (x_j+j-1)}\det_{1\leq j,k\leq l}\left(\frac{1}{(x_j+k-1)!}\right).
\end{align}
Depending on context, either of those three formulae is more convenient. For the imaginary time probabilities, we obtain
\begin{equation}
    P_{x_l,\ldots,x_1}(y,R)=e^{-(R^2-y^2)/4}A_{x_l,\ldots,x_1}\left(\frac{R-y}{2}\right) A_{x_l,\ldots,x_1}\left(\frac{R+y}{2}\right).
\end{equation}
Notice $A_{x_l,\ldots,x_1}$ as given by (\ref{eq:P_det}) is antisymmetric in the $x_1,\ldots,x_l$, which means $P_{x_l,\ldots,x_1}$ is symmetric. One can use this observation to simplify the calculation of the EFP: 
\begin{align}
    E(l,y,R)&=\frac{e^{-\frac{1}{4}r(y)^2}}{l!}\sum_{x_l=-l+1}^\infty\ldots \sum_{x_1=-l+1}^\infty P_{x_l,\ldots,x_1}(y,R)\\\label{eq:gww_before}
    &=\frac{e^{-\frac{1}{4}r(y)^2}}{l!}\oint_{\mathcal{C}^l} \oint_{\mathcal{C}^l}\prod_j \frac{dz_j dw_j e^{\frac{R-y}{2} z_j+\frac{R+y}{2} w_j}}{2\ci \pi(z_j w_j-1)}  \prod_{j<k} (z_j-z_k)(w_j-w_k)\\
    \label{eq:gww_int}
    &=\frac{e^{-\frac{1}{4}r(y)^2}}{l!}\int_{[-\pi,\pi]^l} \prod_j \frac{d\theta_j}{2\pi }e^{\ci(l-1)\theta_j }e^{r(y)\cos\theta_j}\prod_{j<k}2\sin^2\frac{\theta_j-\theta_k}{2}\\\label{eq:gww_det}
    &=e^{-\frac{1}{4}r(y)^2}\det_{0\leq i,j\leq l-1} \left(\int_{-\pi}^{\pi}\frac{d\theta}{2\pi}e^{\ci (j+k-l+1)\theta}e^{r(y)\cos\theta}\right)
\end{align}
where $r(y)=\sqrt{R^2-y^2}$. In going from (\ref{eq:gww_before}) to (\ref{eq:gww_int}), we have deformed the integration contours of the $w_j$ to pick the residues at $w_j=1/z_j$, and then deformed the remaining contours to circles with radius $\sqrt{(R+y)/(R-y)}$. One can recognize in (\ref{eq:gww_int}) the Gross-Witten-Wadia matrix model, for which an exact large deviation is known\cite{GrossWitten,wadia2012}. Set 
\begin{equation}
    X=\frac{l}{R}\qquad,\qquad Y=\frac{y}{R}.
\end{equation}
 We are interested in the limit $R\to\infty$, with $Y\in (-1,1)$ fixed, and $X\in \mathbb{R}_+$ fixed. The asymptotics of the (logarithm of the) EFP can be obtained by using standard minimisation techniques on the log-gas \cite{Forrester} defined by (\ref{eq:gww_int}). The explicit solution reads 
 \begin{equation}\label{eq:largedev}
    \log  E(XR,YR,R)=R^2 \Upsilon(X,\sqrt{1-Y^2})+o(R^2)
 \end{equation}
with rate function \cite{GrossWitten,wadia2012}
\begin{equation}\label{eq:upsilon}
\Upsilon(X,\beta)=\left\{\begin{array}{cc}\frac{\beta^2}{4}-\beta X+\frac{X^2}{4}(3-2\log \frac{X}{\beta})&,\quad X\leq \beta\\
0&,\quad X\geq \beta
\end{array}\right.
\end{equation}
As a function of $X$, $\Upsilon$ decreases from $\beta^2/4$ at $X=0$, to $0$ at $X=\beta$, where it behaves as 
\begin{equation}
\Upsilon(X,\beta)\underset{X\to \beta^-}{\sim}\frac{(\beta-X)^3}{6\beta}.
\end{equation}
The location of the vanishing coincides with the known \cite{prahofer2002scale,Allegra_2016} arctic curve $X^2+Y^2=1$ of the model, see (\ref{eq:largedev}),(\ref{eq:upsilon}) [We will show a more general argument to determine the arctic curve in the next subsection]. The fact that the first and second derivatives vanish at the arctic point is called third order phase transition; it is related to the appearance of the Tracy-Widom distribution for the distribution of the rightmost particle \cite{TracyWidom}. 

We conclude this subsection by pointing out that treating the real time case amounts to straightforward adaptations of the previous calculations. We obtain
\begin{equation}
    E(l,t)=e^{-\frac{t^2}{4}}\det_{0\leq i,j\leq l-1} \left(\int_{-\pi}^{\pi}\frac{d\theta}{2\pi}e^{\ci (j+k-l+1)\theta}e^{t\sin\theta}\right)
\end{equation}
for the real time EFP, which is defined as  $E(l,t)=\braket{\psi(t)|\prod_{j\geq l}\frac{1-\sigma_j^z}{2}|\psi(t)}$. The large $l,t$ the behavior is given by
\begin{equation}
  \log  E(l,t)=-t^2\Upsilon\left(\frac{l}{t},1\right)+o(t^2)
\end{equation}
as $t\to \infty$, with fixed $l/t$, where recall $\Upsilon$ is given by (\ref{eq:upsilon}). Now the vanishing occurs at $l/t=1$, and this coincides with the location of the front in the quench problem \cite{antal1999transport}. As before, the first and second derivatives --but not the third-- vanish at the front, consistent with Tracy-Widom behavior. Let us finally mention that a similar large deviation result for the particle fluctuations in the half space $x\geq 0$ at time $t$ has been found in \cite{Sasamoto_2019}.

\subsection{The trigonometric case $|\Delta|<1$}
\label{sec:trigo}
This parameter range is --except for $\Delta=0$-- by far the most challenging. Let us nevertheless discuss a few  nontrivial applications of our formulae. 
\subsubsection{Asymptotic behavior of $h(\tau|z)$}
First, parametrize $h(\tau|z)$ as follows
\begin{equation}\label{eq:htog}
    h\left(\frac{\tau}{\sin\gamma}\right|\left.\frac{\sin s}{\sin(\gamma+s)}\right)=\frac{\sin s\sin(\gamma+s)}{\sin\gamma} e^{\tau \cot s}g(\tau|s).
\end{equation}
We will need to study the parameter range $z>0$, which imposes $s\in (0,\pi-\gamma)$. 
$g$ satisfies the slightly simpler PDE
\begin{equation}\label{eq:pde_g}
    \left(\tau\partial_\tau^2+\partial_\tau+f(\tau)\right)g(\tau|s)=\partial_s g(\tau|s).
\end{equation}
It is not possible to solve this equation, however for large $\tau$ the Fredholm determinant entering the definition of $f$ --recall (\ref{eq:ortho_f})-- simplifies dramatically \cite{Stephan_return}
\begin{equation}
    \mathcal{Y}(\tau)=\exp\left(\frac{\delta^2}{6}\tau^2+\kappa \log\tau+O(1)\right),
\end{equation}
where
\begin{equation}
    \delta=\frac{\pi}{\pi-\gamma}\qquad,\qquad \kappa=\frac{1}{12}-\frac{\pi}{6\gamma\delta^2}.
\end{equation}
This implies the asymptotic behavior
\begin{equation}
    f(\tau)=\delta^2\tau -\frac{\kappa}{\tau}+o(\tau^{-1})
\end{equation}
which can be inserted in (\ref{eq:pde_g}). To make progress, and inspired by the exact formula at $\gamma=\pi/2$, we make the Ansatz
\begin{equation}
    g(\tau|s)=e^{\tau u(s)}\left(v(s)+\frac{w(s)}{\tau}+O(\tau^{-2})\right)
\end{equation}
and solve the equation order by order. We obtain successively
\begin{align}
    u'&=\delta^2+u^2,\\
    \frac{v'}{v}&=u,\\
    w'&=-(uw+\kappa v),
\end{align}
with solutions
\begin{align}
    u(s)&=\delta \tan (\delta s+A),\\
    v(s)&=\frac{B}{\cos (\delta s+A)},\\
    w(s)&=C\cos (\delta s+A)-\frac{\kappa}{\delta}B\sin (\delta s+A).
\end{align}
Coefficients $A$ and $B$ may be fixed by remembering that $h(\tau|0)=1$ which imposes $g(\tau|s)\sim s^{-1}e^{-\tau/s}$ as $s\to 0$. Hence
\begin{align}
    u(s)&=-\delta \cot \delta s,\\
   v(s) &=\frac{\delta}{\sin \delta s},\\
   w(s)&=C\sin \delta s+\kappa \cos\delta s.
\end{align}
Summing everything up, we obtain
\begin{equation}\label{eq:h_asymtptotics}
    h\left(\frac{\tau}{\sin\gamma}|\frac{\sin s}{\sin(\gamma+s)}\right)=
    e^{\tau(\cot s-\delta \cot \delta s)}
    \frac{\delta\sin s\sin(\gamma+s)}{\sin\delta s \sin\gamma}\left(1+\frac{2C\sin^2 \delta s+\kappa\sin 2\delta s}{2\delta \tau}+\ldots\right)
\end{equation}
where fixing the constant $C$ is not totally obvious. This expansion is needlessly precise, in the following only the leading exponential term will in fact be necessary. The expansion is expected to hold at least for any $s\in (0,\pi-\gamma)$.

\subsubsection{Arctic curves}

\begin{figure}[htbp]
\centering
\begin{tikzpicture}[scale=2]
\draw[ultra thick] (-1.7,-1) -- (1.7,-1);
\draw[ultra thick] (-1.7,1) -- (1.7,1);
\draw[gray,ultra thick] (0,0) circle (1cm);
\draw[red, ultra thick,densely dashed] (0,-1) arc (-90:0:1);
\draw[blue,ultra thick, densely dashed] (1,0) -- (1,1);
\filldraw[red!50!black] (1,0) circle (2pt);
\filldraw[blue!60!black] (1,1) circle (2pt);
\draw (1.2,0) node {$P_1$};
\draw (1.2,0.85) node {$P_2$};
\begin{scope}[xshift=4cm]
\draw[ultra thick] (-1.7,-1) -- (1.7,-1);
\draw[ultra thick] (-1.7,1) -- (1.7,1);
\draw[gray,ultra thick] (0,0) circle (1cm);
\draw[red, ultra thick,densely dashed] (0,-1) arc (-90:-20:1);
\draw[blue,ultra thick, densely dashed] (0.94,-0.342) -- (1.43,1);
\filldraw[red!50!black] (0.94,-0.342) circle (2pt);
\filldraw[blue!60!black] (1.43,1) circle (2pt);
\draw (1.15,-0.342) node {$P'_1$};
\draw (1.6,0.85) node {$P'_2$};
\end{scope}
\begin{scope}[yshift=2.3cm]
 \node (myfirstpic) at (0,0) {\scalebox{1}[-1]{\includegraphics[width=6cm]{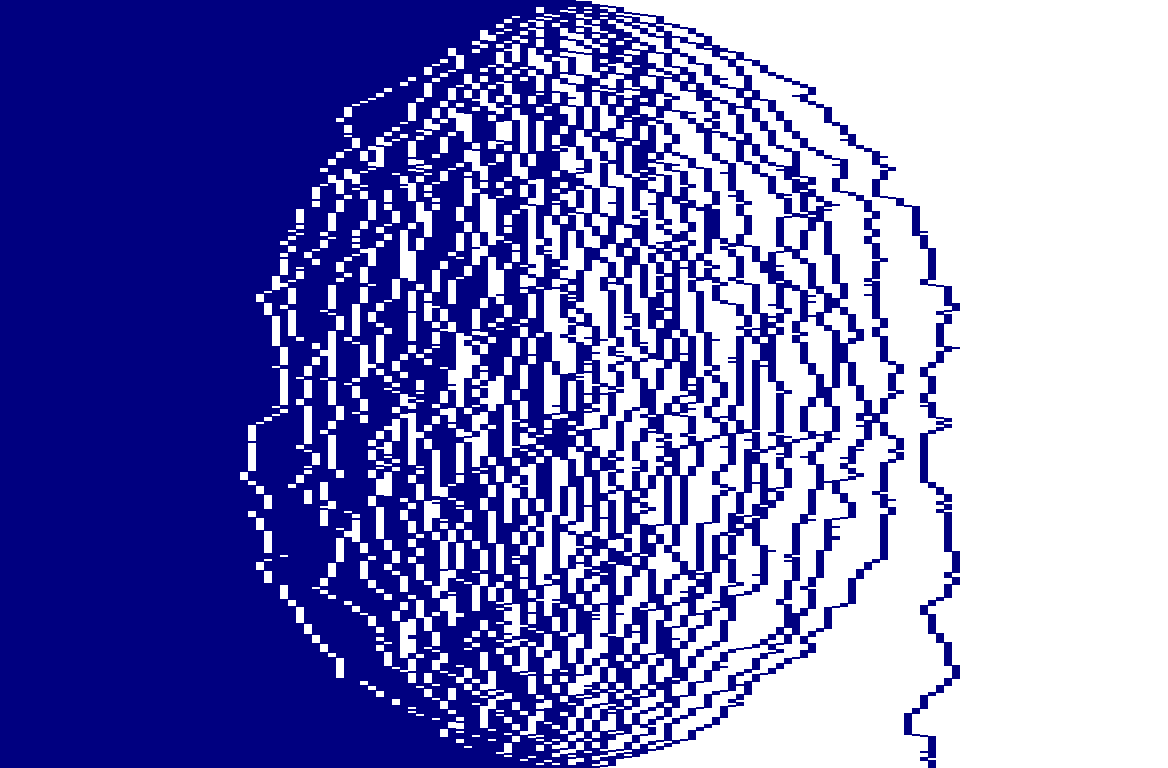}}};
\end{scope}
\end{tikzpicture}
\caption{Illustration of the tangent method \cite{ColomoSportiello_tangent} used to determine the arctic curve in the XXZ spin chain. Top left: typical configuration, with one particle not conditioned to go back to its initial location. Bottom left: most likely trajectory for the test particle in the scaling limit. It follows the arctic curve until point $P_1$, then goes  straight ahead until reaching the upper boundary at point $P_2$. Bottom right: most likely trajectory if we enforce the test particle to end at point $P'_2\neq P_2$. In that case, the particle follows the arctic curve until point $P'_1$, leaves it  tangentially and connects to $P'_2$ through a straight line.}
\label{fig:tangent}
\end{figure}
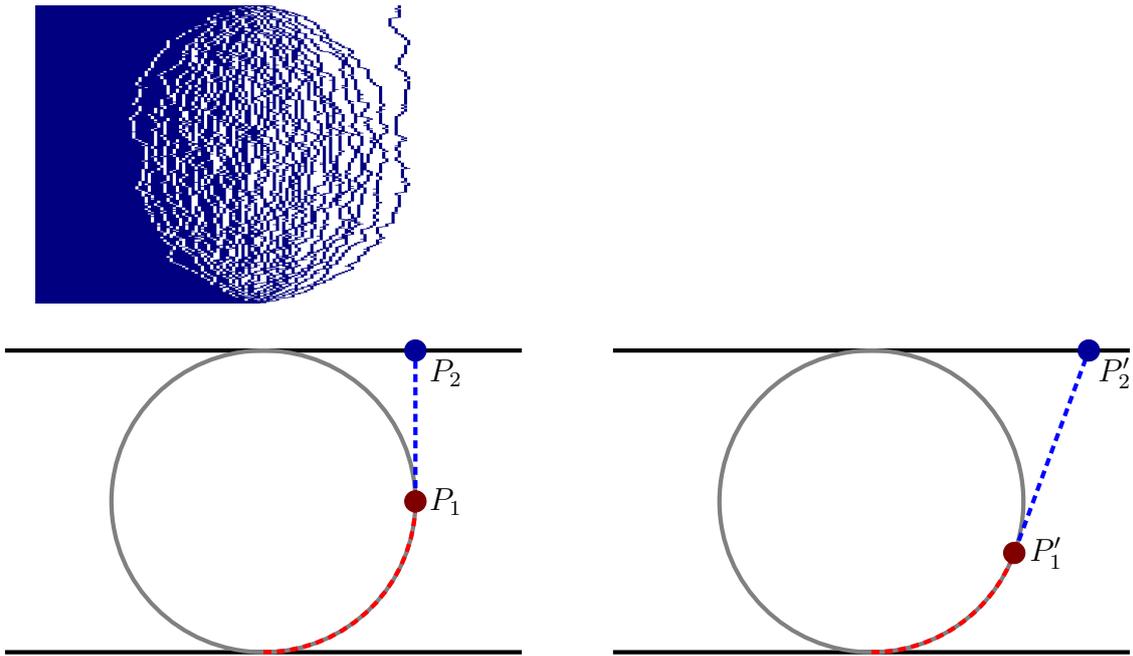

In the following, we will discuss an application of the dominant term in the expansion of $h(\tau|z)$, $h(\tau|z)\sim e^{\tau F(z)}$ and show that this term alone allows to reconstruct the arctic curve, which is already a highly nontrivial result. The first step is to evaluate the quantity 
\begin{equation}
\mathcal{A}(X)=\lim_{R\to\infty} \frac{\log A_{XR}(R)}{R}
\end{equation} 
for $X\geq 0$. 
Because of (\ref{eq:hdef}), $A_{XR}(R)$ can be written as a contour integral, which reads
\begin{equation}
    A_{XR}(R)= \oint_{\mathcal{C}} \frac{dz}{2\ci \pi z} e^{R[ F(z)-X\log z]}.
\end{equation}
A saddle point treatment for large $R$ is straightforward. We obtain
\begin{equation}
    \mathcal{A}(X)=F(z_s)-X\log z_s,
\end{equation}
where $z_s$ is the solution of
\begin{equation}\label{eq:nonlineareq}
    z_s F'(z_s)=X.
\end{equation}
It is easy to find the location $X_0$ with maximal amplitude $\mathcal{A}(X_0)$. With $z_s$ solution of (\ref{eq:nonlineareq}), we obtain $\frac{d\mathcal{A}}{dX}=-\log z_s=0$ iff $z_s=1$, which means
\begin{align}\label{eq:exact_arctic0}
    X_0&=F'(1)
=-1+\delta^2 \cos^2 \frac{\gamma}{2}.
\end{align}
To get the last equality, we have used
\begin{equation}
    F\left(\frac{\sin s}{\sin (\gamma+s)}\right)=\cot s-\delta \cot \delta s,
\end{equation}
see (\ref{eq:h_asymtptotics}). 
We also have
\begin{equation}\label{eq:free_x0}
    F(X_0)=1-\Delta.
\end{equation}
This result has a nice interpretation, which is pictured in figure \ref{fig:tangent} (left).
The logic is that for large $\tau$, there are of order $\tau$ particles which start moving to the right at imaginary time $0$. Those start coming back to their initial positions at imaginary time $\tau/2$, before reaching them at imaginary time $\tau$.  The only particle which is not conditioned to go back to its initial place is  unlikely to influence the many others, but the reverse is of course not true. The analogous interpretation in the slab geometry $y\in(-R,R)$ with Hamiltonian (\ref{eq:newH}) follows from a simple affine transformation of the above, and this is the point of view we take in the following. 

The Hamiltonian contains both left and right hopping terms, so in the absence of the others the particle wants to go straight on average.
For imaginary time $y\in[-R,0]$ this is not possible, since the other particles push it do the right. The next best thing is to follow the arctic curve until the particles are furthest away to the right at $y=0$. After that the particles start going back to their initial position, leaving the test particle free from their influence. Therefore it leaves the arctic curve at point $P_1$ tangentially, hitting point $P_2$ at imaginary time $R$ (see figure \ref{fig:tangent} left). By this logic, $P_2$ must correspond to the point of maximal amplitude $X_0$. $P_1$ is the arctic point corresponding to $Y=0$, so (\ref{eq:exact_arctic0}) provides an exact formula for the location of this point since $P_1$ and $P_2$ have the same abscissa. Note $X_0=1$ for $\Delta=0$, consistent with the results of the previous subsection.

Before pushing the interpretation further, let us pause and make more quantitative statements about the propagation of a free particle in the vacuum $\ket{0}=\ket{\ldots\downarrow\downarrow\downarrow\ldots}$. The single particle propagator can be computed exactly as follows:
\begin{align}
    C_j(\tau)&=\braket{0|\sigma_j^- e^{\tau H_{XXZ}}\sigma_0^+|0},\\
    &=e^{-\Delta\tau}\int_{-\pi}^{\pi} \frac{d\theta}{2\pi} e^{\ci j \theta+\tau\cos\theta},
\end{align}
which can also be written as a modified Bessel function. Saddle point analysis yields
\begin{equation}\label{eq:C_asymptotics}
    C_{X\tau }(\tau)=\exp\left(\tau g(X)+O(\log \tau)\right)
\end{equation}
for large $\tau$, 
with
\begin{equation}\label{eq:functiong}
    g(X)=\sqrt{1+X^2}-X\log(X+\sqrt{1+X^2})-\Delta.
\end{equation}
$g$ is maximal at $X=0$, consistent with the fact that the particle wants to go straight. If we condition the particle to end at a different position $X\tau$, the most likely trajectory will still be a straight line, as essentially follows from the concavity of $g$\footnote{This can be shown by considering $\braket{0|\sigma_{X\tau}^-  e^{\tau(1-\omega)H}\sigma_{X'\tau}^+ \sigma_{X'\tau}^- e^{\omega\tau}\sigma_{0}^+ |0}=I_{X'\tau}(\omega\tau)I_{(X-X')\tau}(\tau(1-\omega))$ with   $\omega\in (0,1)$. For large $\tau$, this is maximal at position $X'=\omega X$, as follows from the expansion (\ref{eq:C_asymptotics}) and the concavity of $g$. In going from position $(0,0)$ to position $(X\tau,\tau)$, the particle most likely passed through point $(X\omega\tau,\omega\tau)$ for any $\omega$, implying a straight line is far more likely than any other trajectory.}. Assuming separation of scales, one can also use (\ref{eq:C_asymptotics}) to compute the propagator associated to any sufficiently smooth trajectory connecting two well separated points.

One can generalize the argument leading to the determination of the arctic point $P_1$ to reconstruct the full arctic curve. The general procedure is known as the tangent method \cite{ColomoSportiello_tangent}; it has been shown to apply to a wide class of --interacting or not-- particle systems, see e.g. \cite{Di_Francesco_2018,Debin_2019,DebinDiFrancescoGuitter,Aggarwal2020}. As we shall see below, the only ingredient needed will be the asymptotic behavior of $h(\tau|z)$, combined with the free propagation result (\ref{eq:C_asymptotics},\ref{eq:functiong}). 

Assume now that the region inside the arctic curve is convex. We parametrize the right part of the curve as $(X(s),Y(s))$ for $s\in [0,\pi-\gamma]$ (this choice of parametrization is purely for later convenience). $Y$ is an increasing function with boundary values $Y(0)=-1$, $Y(\pi-\gamma)=1$. $X$ satisfies $X(0)=0$, increases with $Y$ for $Y\leq 0$, reaches its maximum $X_0$ at $Y=0$, and decreases back to $X(\pi-\gamma)=0$ afterwards. Now force the particle to end at a point $P'_2$, different from the most likely escape point $P_2$, as shown in figure \ref{fig:tangent} (right). What is the most likely trajectory? Clearly, the test particle has to follow the arctic curve for a while, before leaving the curve at some point $P'_1$ and connecting to $P'_2$ through a straight line. For small $s$ this is not possible, since the straight line would have to go through the bulk, which is hugely disfavoured. The first point for which this can occur is the point $P'_1$ such that segment $[P'_1,P'_2]$ is tangent to the arctic curve. Since a straight line is always favoured over more complicated trajectories, the most likely trajectory is therefore to follow the arctic curve until the tangencency point $P'_1$, and then to  follow the segment $[P'_1,P'_2]$. 

The previous argument can be made quantitative. The tangent to the curve at "time" $s$ has equation $Y-Y(s)=\frac{Y'(s)}{X'(s)}(X-X(s))$. Denote by $(X_2(s),1)$ the point where the test particle hits the top boundary. From the previous equation
\begin{equation}
    X_2(s)=X(s)+\left(1-Y(s)\right)\frac{X'(s)}{Y'(s)},
\end{equation}
and the tangencency assumption implies
\begin{equation}\label{eq:tangent_constraint}
    \mathcal{A}(X_2(s))=\int_{0}^{s} Y'(u) g\left(\frac{X'(u)}{Y'(u)}\right)\,du \;+\;(1-Y(s))g\left(\frac{X'(s)}{Y'(s)}\right).
\end{equation}
On the rhs of the previous equation, the first term corresponds to the energy associated to following the arctic curve until the tangencency point, the second term is the contribution of the straight line segment. Equating this to the exact free energy $\mathcal{A}$ obtained by exact methods --see previous subsection-- is sufficient to uniquely determine the arctic curve. Before doing that, it is worth noting that the rhs is determined from purely free particle arguments, but the lhs, in contrast, does depend on interactions. The fact that the rightmost particle is also free while following the arctic curve can be justified by a dilution argument, see e.g. \cite{Stephan_free_edge}.

To solve (\ref{eq:tangent_constraint}), it is convenient to take time derivative on both sides. We obtain
\begin{equation}\label{eq:tangent_constraint2}
    \log z_s(X_2(t))=\textrm{arcsinh}\, \frac{X'(t)}{Y'(t)}
\end{equation}
where recall $z_s(X)$ is the real solution to the equation $z_s F'(z_s)=X$. The equation (\ref{eq:tangent_constraint}) does admit a somewhat complicated by fully explicit solution. Introducing the ancillary functions
\begin{align}
    \Phi(s)&=\frac{\sin\gamma}{\sin s\sin(\gamma+s)}\\
    \chi(s)&=\cot(\gamma+s)-\delta\cot\delta s,
\end{align}
one can check that 
\begin{align}
    X(s)&=\sin\gamma \frac{\Phi'(s) \chi''(s)-\chi'(s)\Phi''(s)}{\Phi'^2(s)-\Phi(s)\Phi''(s)}
    \\
    Y(s)&=1+2 \frac{\Phi'(s)\chi'(s)-\Phi(s)\chi''(s)}{\Phi'^2(s)-\Phi(s)\Phi''(s)}
\end{align}
is a solution to (\ref{eq:tangent_constraint2}). This solution is in exact agreement with the result of \cite{Stephan_return}, obtained using different methods. One can check that this arctic curve is a circle at the free fermions point ($\gamma=\pi/2$), an algebraic curve at roots of unity, non algebraic otherwise. Interestingly, the inverse slope of the tangent at point $s$ is given by the simpler looking formula
\begin{equation}\label{eq:slope}
    \frac{X'(s)}{X(s)}=-\frac{\sin\gamma}{2}\, \frac{\Phi'(s)}{\Phi(s)},
\end{equation}
while the abscissa of the escape point $P_2'$ is
\begin{equation}\label{eq:escape}
    X_2(s)= \sin\gamma\, \frac{\chi'(s)}{\Phi(s)}.
\end{equation}
For any $s$, the knowledge of the escape point and the slope of the tangent specifies a unique straight line. This means equations (\ref{eq:slope}), (\ref{eq:escape}) define a family of straight lines parametrized by $s$. The arctic curve can then simply be interpreted\cite{ColomoSportiello_tangent} as the envelope of this family of straight lines, similar to a caustic in ray optics.
 
 Let us finally comment on the previous prediction  \cite{Stephan_return} of the arctic curves, which was done by crudely taking the Hamiltonian limit of the arctic curves in the six vertex model \cite{ColomoPronkocurve}. The present derivation is of course still heuristic but nevertheless more satisfying, since we did not need to assume that the scaling limit and the Hamiltonian limit commute. One conclusion from our study is therefore that commuting the two is physically harmless in imaginary time. Unsurprisingly, this claim also hold for the asymptotics of $h(\tau|z)$ which can also be obtained from taking the Hamiltonian limit of the asymptotics \cite{ColomoPronkocurve} of the analogous $h_N(\epsilon|z)$ in the six vertex model. 
\subsubsection{Emptiness formation probability}

Our final example is an exact formula for the emptiness formation probability discussed in sections \ref{sec:gapped} and \ref{sec:freefermions}, which holds for arbitrary values of $\Delta$. With the two main formulae (\ref{eq:multiamplitude_bis}) and (\ref{eq:pde}) at hand, the result becomes a trivial consequence of the main result of Ref.~\cite{ColomoDiGiulioPronko}. We report the formula here for completeness:
\begin{equation}\label{eq:efp_delta_imag}
    \!\!\! \frac{l!E_l}{E_0}=\!\oint_{\mathcal{C}^l} \prod_{j=1}^l \frac{dz_j}{2\ci\pi z_j}h\!\left(\frac{R-y}{2}\right|\!\left.\frac{1}{z_1},\ldots,\frac{1}{z_l}\right)\!h\!\left.\left(\frac{R+y}{2}\right|z_1,\ldots,z_l\right)\!\!\prod_{1\leq j\neq k\leq l} \frac{z_k-z_j}{z_j z_k-2\Delta z_j+1}
\end{equation}
where we have used the shorthand notation $E_l$ instead of $E(l,y,R)$.  It would be tempting to study the large time asymptotics by using the same strategy as in the previous subsections, namely using (\ref{eq:h_asymtptotics}), which is of the form (\ref{eq:h_ideal}) to leading order, meaning (\ref{eq:multiamplitudes_ideal}) holds to leading order. In that case one would get a matrix integral generalizing the Gross-Witten-Wadia one at $\Delta=0$, (see (\ref{eq:gww_int})). To perform this analysis, one would need a more precise asymptotic estimate which would also hold in some large enough region of the complex plane (for example at $y=0$ it is very natural to integrate on unit circles).  
Our estimate was derived assuming real $z>0$, even though it might extend further. We expect changes in asymptotics to occur at least at $z=e^{\pm\ci\gamma}$, since some terms in (\ref{eq:h_asymtptotics}) vanish in that case. 
Also, it is not completely clear up to which values of $l$ the estimate can be safely used. We leave this as an interesting open problem, possibly more of an acid test of the approach advocated here, similar to difficulties encountered in the six vertex model \cite{ColomoDiGiulioPronko}. 

To obtain the real time result, just make the replacements $y\to \ci t$ and $R= 0$ in the previous formula. One may choose the integration contours to be unit circles, in which case the formula reads
\begin{equation}\label{eq:efp_delta_real}
    \frac{l!\mathcal{F}_l}{\mathcal{F}_0}=\int_{[-\pi,\pi]^l} \prod_{j=1}^l \frac{d\theta_j}{2\pi} \left|h\left(\ci t\left|e^{\ci\theta_1},\ldots,e^{\ci\theta_l}\right)\right.\prod_{1\leq j<k\leq l} \frac{e^{\ci \theta_k}-e^{\ci\theta_j}}{e^{\ci (\theta_j+\theta_k)}-2\Delta e^{\ci \theta_j}+1}\right|^2
\end{equation}
An asymptotic analysis of this multiple integral is well outside the scope of the present paper. 
\subsection{The Heisenberg point $\Delta=1$}
\label{sec:heisenberg}
Our last example is the rational point $\Delta=1$, which is at the boundary between the two very different regimes studied in section  \ref{sec:gapped} and section \ref{sec:trigo}, and does not have a physical analog on the six vertex side. As is known, the Hamiltonian at $\Delta=1$ coincides exactly with the generator of the symmetric simple exclusion (stochastic) process (SSEP) \cite{Sandow1994}. Initialize now SSEP in a state with all particles packed to the left of the origin. Our amplitudes give access to the probability that the $l$ rightmost particles end up at positions $x_1,\ldots,x_l$, conditioning all the others on coming back to their initial location. 

SSEP is a well known example of a diffusive process \cite{GwaSpohn,Golinelli_2006,Lazarescu_2015} described by macroscopic fluctuation theory \cite{MFT}, so let us try and recover some of this behavior from our exact solution. The partition function has been studied in \cite{Stephan_return} (see also \cite{DerridaGerschenfeld}), which derived the large $\tau$ expansion
\begin{equation}
    \log Z(\tau)=-\zeta(3/2)\sqrt{\frac{2\tau}{\pi}}+\frac{1}{4}\log\tau +O(1).
\end{equation}
$\zeta(u)$ is the Riemann Zeta function $\zeta(u)=\sum_{n\geq 1} n^{-u}$. 
From this we can deduce the large $\tau$ behavior of the ancillary function
\begin{equation}
    Q(\tau)=-\frac{1}{4\tau}+O(\tau^{-3/2}),
\end{equation}
and plugging the leading term in the PDE (\ref{eq:pde}) yields
\begin{equation}\label{eq:pde_diffusive}
   \left(\tau\partial_\tau^2+\left[1-2\tau \left(\frac{1}{z}-1\right)\right]\partial_\tau-\frac{1}{4\tau}-z+1\right)h(\tau|z)=(1-z)^2\partial_z h(\tau|z)
\end{equation}
Now one of the simplest argument to show diffusion goes as follows. 
For $z$ sufficiently small, the $\tau$ dependence disappears for large $\tau$, meaning $(1-z)\partial_z h(\tau|z)=h(\tau|z)$ so \begin{equation}
h(\tau|z)\sim \frac{1}{1-z}.
\end{equation}
This implies 
\begin{equation}\label{eq:one}
    A_x(\tau)\sim 1
\end{equation} for finite $x$ and large $\tau$. Now observe that the equation (\ref{eq:pde_diffusive}) at $z=1$ becomes very simple
\begin{equation}
   \left( \tau\partial_\tau^2+\partial_\tau-\frac{1}{4\tau}\right)h(\tau|1)=0.
\end{equation}
It has an exact solution $h(\tau|1)=\alpha_1\sqrt{\tau}+\alpha_2/\sqrt{\tau}$ for unknown $\alpha_1,\alpha_2$. Therefore $h(\tau|1)=\sum_{x\geq 0}A_x(\tau)$ is of order $\sqrt{\tau}$ for large $\tau$, which combined with (\ref{eq:one}) strongly suggests diffusive behavior, as should be. We leave a more precise treatment for future work. Since the real time asymptotics of the return amplitude are also known \cite{Stephan_return}, it is reasonable to expect similar behavior in the real time problem also. This is discussed further in the conclusion.

\section{Conclusion}
In this paper, we have derived some exact formulae for the (imaginary or real) time evolution of the XXZ spin chain starting from a domain wall initial state. These formulae were obtained as a limit of the six vertex model, using the approach of \cite{Stephan_return} and combining with results from the theory of orthogonal polynomials. We also checked these to high order using a power series method. 

From these formulae, it is in principle possible to reconstruct all time evolved local observables and correlations. This task is still a very  difficult one, as is typical for interacting integrable systems. Nevertheless, we demonstrated on a few very simple observables how this can be (fully or partially) be done, for various values of the interaction parameter $\Delta$. In particular, the generating function $h(\tau|z)$ for the 'single particle' amplitude already contains non-trivial physical information, as it allows to derive the arctic curves by making use of the  tangent method. 

There are many directions in which our results could be expanded. First, our asymptotic results for $h(\tau|z)$ are somewhat primitive, as they hold only for $z>0$. They were obtained by assuming a given asymptotic form --similar to the six vertex one\cite{ColomoPronkocurve}-- and fixing the parameters/functions by consistency with respect to the PDE. This turned out to be sufficient for our purposes, but it would be interesting to extend the asymptotic results to the whole complex plane; this would be necessary to study more complicated observables, as can already be seen from the exact formula for the EFP (\ref{eq:efp_delta_imag}). One can check that the asymptotic expansion vanishes for certain values of $z$, which hints at a change in asymptotics. This guess is also supported by numerical simulations not shown here.

Most of our illustrative examples were dealing with imaginary time, but it would be very satisfying to be able to make contact with the exact results from GHD  by taking the $t\to \infty$ limit in our approach. In particular, the fact that the hydrodynamic  density profile is given by an extremely simple formula\cite{ColluraDeLucaViti} does suggest that it is possible. One could try to use a similar strategy as we did here, by boldly plugging the asymptotic expansion of the return amplitude in the main PDE, and see if one can get meaningful results out of it. For the return amplitude, it was conjectured in \cite{Stephan_return} that
\begin{align}
    \log |Z(\ci t)|&\sim \frac{q^2}{q^2-1}\frac{(t\sin\gamma)^2}{12}\qquad,\qquad\pi\gamma=\frac{p}{q} \in\mathbb{Q}
    \\
    &\sim t\sin\gamma \qquad\qquad \qquad,\qquad\pi\gamma\notin \mathbb{Q}
\end{align}
The conjecture is a guess (educated on exact imaginary time asymptotics) which was checked using on high precision numerics on the Fredholm determinant. The asymptotic form shows singular behavior with $\Delta$, which is also known to occur in the TBA treatment \cite{takahashi_1999} of the domain wall state, where only two string solutions turn out to play a role, similar to the famous Drude weight problem\cite{Xotos,ProsenIlievski_drude,IlievskiDeNardis,ColluraDeLucaViti} in the spin chain. It would be interesting to see how much information may be extracted by this crude replacement, and how to generalize this idea to recover GHD predictions if possible.  

Another interesting project would be to make contact with exact results based on finite particle Bethe Ansatz (see \cite{FeherPozsgay} for the XXZ chain, \cite{Tracy2008,Tracy2008bis} for the asymmetric exclusion process, and \cite{DerridaGerschenfeld}, for SSEP). The analogous real-time problem at $\Delta=1$ is also very interesting, with connections to a classical Landau-Lifshitz equation \cite{GamayunMiaoIlievski,Misguich_LL,Bulchandani_2021} on the hydrodynamic side. This effective model predicts logarithmically enhanced diffusion, consistent with the best numerical results so far \cite{Misguich_numerics}. The distribution of the rightmost particle has also been studied numerically \cite{Stephan_free_edge} in that model, it would be also interesting to investigate it using the exact approach advocated here.

Finally, one could also try to generalize slightly the setup studied in the present paper. For example, it should be possible to add other conserved charges to the XXZ Hamiltonian, as was done in \cite{Bocini_2021} at the free fermion point. In this case the imaginary time model is not even positive, which leads to intriguing behavior for e.g. the density. Another interesting generalization would be too study other simple initial states such as the mixed state studied in \cite{YoungItalians,Ljubotina2017}. Our approach clearly relied on a somewhat miraculous connection to the six vertex model with domain wall boundary conditions, but it could be that other such states can be handled starting from the six vertex model also. In that case, one could (at $\Delta=1$) try to make contact with the literature on superdiffusion in spin chains with additional symmetries (see \cite{Bulchandani_2021} and references therein).     

\label{sec:conclusion}
\ack
I wish to thank Jérémie Bouttier, Filippo Colomo, Benoit Estienne, Christian Hagendorf, Karol Kozlowski, Alexandre Lazarescu, Vincent Pasquier, Lorenzo Piroli, Tomohiro Sasamoto and Eric Vernier for discussions at various stages of the project. This work was partially supported by the ANR-18-CE40-0033 grant 'Dimers'.
\pagebreak

\appendix
\section[\qquad\qquad\quad A limit of orthogonal polynomials]{A limit of orthogonal polynomials}
\label{app:orthopolys}
Let us consider a weight function $w(x)$ on $\mathbb{R}$ which induces a scalar product
\begin{align}
    \braket{f|g}=\int_{\mathbb{R}}dx\, f(x)g(x)w(x).
\end{align}
A classical analysis problem deals with the construction of polynomials $p_n$ of degree $n$, which are orthonormal with respect to this scalar product:
\begin{equation}
    \braket{p_n|p_m}=\delta_{nm}.
\end{equation}
In this appendix, we are interested in weight functions of the form
\begin{equation}
    w_\epsilon(x)=e^{-\epsilon x}w_0(x)
\end{equation}
for some $\epsilon>0$. $w_0$ is a  nonnegative function which satisfies the two properties:
\begin{enumerate}
    \item $w_0(x)\to 1$ sufficiently fast as $x\to\infty$.
    \item $w_\epsilon(x)\to 0$ sufficiently fast as $x\to-\infty$, uniformly in $\epsilon$. 
\end{enumerate}
We denote by $p_n$ the corresponding orthonormal polynomials, or $p(n,\epsilon|x)$ to emphasize the dependence on $\epsilon$. The sign ambiguity is lifted by assuming $p(n,\epsilon|0)>0$ \footnote{This is a non standard convention, which will however be convenient for our purposes.}. In this appendix, we will study the limit
\begin{equation}\label{eq:ortholimit}
    q(\alpha|x)=\lim_{n\to\infty}\sqrt{\frac{n}{\alpha}} \,p(n,\alpha/n|x),
\end{equation}
and show that it satisfies the differential equation
\begin{equation}\label{eq:ortho_ode}
    \left[\alpha\partial_\alpha^2+\partial_\alpha+f(\alpha)+x\right]q(\alpha|x)=0,
\end{equation}
where
\begin{equation}\label{eq:ortho_f}
    f(\alpha)=2\alpha \frac{d^2\log \mathcal{Y}(\alpha)}{d\alpha^2}+\frac{d\log \mathcal{Y}(\alpha)}{d\alpha},
\end{equation}
and $\mathcal{Y}(\alpha)$ is the Fredholm determinant
\begin{equation}\label{eq:ortho_det}
    \mathcal{Y}(\alpha)=\det(I-V),
\end{equation}
with kernel
\begin{equation}\label{eq:ortho_kernel}
    V(x,y)=\frac{\sqrt{y}J_0(2\sqrt{x})J_0'(2\sqrt{y})-\sqrt{x}J_0(2\sqrt{y})J_0'(2\sqrt{x})}{(x-y)} \left[\Theta(y)-w_0\left(\frac{y}{\alpha}\right)\right]
\end{equation}
acting on functions in $L^2(\mathbb{R})$. We use the definition
\begin{equation}\label{eq:fred_def}
    \det(I-V)=\exp\left(-\sum_{n\geq 1}\frac{1}{n}\int_{\mathbb{R}^n}dx_1,\ldots dx_n V(x_1,x_2)\ldots V(x_{n-1},x_n)V(x_n,x_1)\right)
\end{equation}
for the Fredholm determinant. 
The appendix is organized as follows. In \ref{app:examples} we take equations (\ref{eq:ortholimit},\ref{eq:ortho_ode},\ref{eq:ortho_f},\ref{eq:ortho_det},\ref{eq:ortho_kernel}) for granted, and discuss simple examples. We then recall some basic orthogonal polynomial theory in \ref{app:ortho_elementary}, before proceeding with the actual derivation in \ref{app:ortho_derivation}, \ref{app:ortho_derivation2}, \ref{app:ortho_derivation3}.
\subsection[\qquad\qquad\quad Examples]{Examples}
\label{app:examples}
The simplest example one can think of is the case $w_0(x)=\Theta(x)$, that is a weight $w_\epsilon(x)=e^{-\epsilon x}$ on $\mathbb{R}_{+}$. The corresponding orthonormal polynomials are
\begin{equation}
    p(n,\epsilon,x)=\sqrt{\epsilon}L_n(\epsilon x),
\end{equation}
where 
\begin{equation}\label{eq:laguerrepoly}
L_n(x)=\sum_{k=0}^n C_n^k (-x)^k/k!
\end{equation}
are the Laguerre polynomials. Taking the limit is straightforward, and we obtain
\begin{equation}\label{eq:example_bessel}
    q(\alpha|x)=J_0(2\sqrt{\alpha x}).
\end{equation}
The limit is often used in Random matrix theory \cite{Forrester}, where of course the differential equation is not needed. In this example $\mathcal{Y}(\alpha)=1$ trivially, so  $f(\alpha)=0$ and one can check that (\ref{eq:example_bessel}) solves the ODE (\ref{eq:ortho_ode}) with $f=0$.

Another simple example is a particular case of the Meixner-Pollaczek weight
\begin{equation}
    w_0(x)=\frac{1}{1+e^{-\pi x/2}}.
\end{equation}
The orthonormal polynomials are given explicitely \cite{ColomoPronkodet} in terms of the Gauss hypergeometric function
\begin{equation}\label{eq:mp_polynomials}
    p(n,\epsilon|x)=e^{2\ci\epsilon n}\sqrt{\sin\epsilon\cos\epsilon} \, _{2} F_1(-n,1/2+\ci x/4;1|1-e^{-4\ci \epsilon}).
\end{equation}
Taking the limit, we obtain
\begin{equation}\label{eq:example_1f1}
    q(\alpha|x)=e^{2\ci \alpha}\,_1F_1(1/2-\ci x/4;1|-4\ci \alpha).
\end{equation}
The Fredholm determinant can also be computed explicitly, for example by using the polynomials (\ref{eq:mp_polynomials}) and taking the limit. We get successively $\mathcal{Y}(\alpha)=e^{2\alpha^2/3}$ and $f(\alpha)=4\alpha$. One can also check that (\ref{eq:example_1f1}) solves the ODE (\ref{eq:ortho_ode}) for this choice of $f$.

The example relevant to the present paper is the weight
\begin{equation}
    \label{eq:example_6v}
    w_0(x)=\frac{1-e^{-\gamma x}}{1-e^{-\pi x}}
\end{equation}
for $\gamma\in(0,\pi)$, see (\ref{eq:weight_fct}) in the main text. 
The previous one corresponds to the particular case $\gamma=\pi/2$ (free fermions, $\Delta=0$), but away from this point the orthonormal polynomials are not known  explicitly. Neither is the Fredholm determinant (\ref{eq:ortho_det}), which means there is, as expected, no simple explicit solution for $q(\alpha|x)$. The ODE still proves useful, as demonstrated in section \ref{sec:examples}.
\subsection[\qquad\qquad \quad Reminder on orthogonal polynomials theory]{Reminder on orthogonal polynomials theory}
\label{app:ortho_elementary}
Let us introduce the determinant
\begin{equation}
    D_n=\det_{0\leq i,j\leq n-1}\left(\braket{x^{i+j}}\right)
\end{equation}
where $\braket{x^{i+j}}=\int_{\mathbb{R}} x^{i+j}w(x)\,dx$, and its cousin
\begin{equation}
    D'_n=\det_{0\leq i,j\leq n-1}\left(x^{i+j+\delta_{i,n-1}}\right),
\end{equation}
which differs from the previous one only in the last row. The orthonormal polynomials can be expressed as a $(n+1)\times (n+1)$ determinant
\begin{equation}\label{eq:orthodet}
    p_n(x)=\frac{(-1)^n}{\sqrt{D_n D_{n+1}}}\left|\begin{array}{ccccc}
    \braket{x^0}&\braket{x^1}&\ldots& \braket{x^{n}}\\
    \vdots&&&\vdots\\
    \braket{x^{n-1}}&\braket{x^{n}}&\ldots&\braket{x^{2n-1}}\\
    1&x&\ldots&x^n
    \end{array}
    \right|.
\end{equation}
A fundamental role is played by the three-term recurrence \cite{Szego}
\begin{equation}\label{eq:threeterms}
    xp_n(x)=-a_{n} p_{n+1}(x)+b_n p_n(x)-c_{n} p_{n-1}(x),
\end{equation}
where
\begin{align}\label{eq:an}
a_n&=\sqrt{\frac{D_n D_{n+2}}{D_{n+1}^2}},\\\label{eq:bn}
b_n&=\frac{D_{n+1}'}{D_{n+1}}-\frac{D_n'}{D_n},\\\label{eq:cn}
c_n&=a_{n-1}.
\end{align}
As we shall see, taking the limit (\ref{eq:ortholimit}) in this recurrence relation ultimately gives the ODE (\ref{eq:ortho_ode}).
The orthonormal polynomials can also be used to compute the determinant and inverse of Hankel matrices $A_{ij}=(\braket{x^{i+j}})_{i,j}$, where $i,j\in\{0,\ldots,n-1\}$. Denote by $k_n$ the coefficient of the leading term in $p_n$. The determinant is
\begin{equation}
    \det A=\prod_{p=0}^{n-1} \frac{1}{k_p^2} ,
\end{equation}
while the inverse reads
\begin{equation}\label{eq:inverse}
    (A^{-1})_{ij}=\left.\frac{1}{i!j!}\frac{\partial^{i+j}K_n(x,y)}{\partial x^i \partial y^j} \right|_{x=y=0},
\end{equation}
where
\begin{equation}\label{eq:cdformula}
    K_n(x,y)=-\frac{k_{n-1}}{k_n}\frac{p_n(x)p_{n-1}(y)-p_{n-1}(x)p_n(y)}{x-y}
\end{equation}
$K_n$ is called the Christoffel-Darboux kernel.
\subsection[\qquad\qquad\quad The three term recurrence relation]{The three term recurrence relation}
\label{app:ortho_derivation}
Our starting point is the three-term recurrence relation (\ref{eq:threeterms}) which we rewrite as
\begin{equation}\label{eq:threeterms_anoother}
    xp_n=\left(b_n-a_n-c_n\right)p_n+\left(c_n-a_n\right)\frac{p_{n+1}-p_{n-1}}{2}-(c_n+a_n)\frac{p_{n+1}-2p_n+p_{n-1}}{2}.
\end{equation}
We assume the form
\begin{equation}
    p(n,\epsilon|x)\sim \sqrt{\epsilon}\sum_{k=0}^n n^k q_k(x) \epsilon^k
\end{equation}
holds for large $n$ and small $\epsilon$. $q_k$ is expected to be a polynomial in $x$ (by the arguments in section \ref{sec:introduction}, see also \ref{app:checks}). The previous assumption implies $q(\alpha|x)=\sum_{k=0}^\infty q_k(x)\alpha^k$. Using this logic, we obtain
\begin{align}
    \frac{p(n+1,\alpha/n,x)-p(n-1,\alpha/n,x)}{2}&\sim
    \sqrt{\alpha/n} \sum_{k=0}^n q_k(x) \frac{(n+1)^k-(n-1)^k}{2}(\alpha/n)^k\\
    &\sim  \sqrt{\alpha/n} \frac{1}{n}\sum_{k=0}^\infty k q_k(x)\alpha^k\\
    &\sim \sqrt{\alpha/n} \frac{1}{n}\, \alpha\partial_\alpha q(\alpha|x).
\end{align}
By a similar argument,
\begin{align}
   \frac{p(n+1,\alpha/n,x)-2p(n,\alpha/n,x)+p(n-1,\alpha/n,x)}{2}\sim \sqrt{\alpha/n} \frac{1}{2n^2}\,\alpha^2 \partial_\alpha^2 q(\alpha|x),
\end{align}
so the recurrence turns into
\begin{equation}\label{eq:crucial_step}
    x q(\alpha|x)\sim (b_n-a_n-c_n)q(\alpha|x)+\frac{c_n-a_n}{n}\alpha\partial_\alpha q(\alpha|x)-\frac{c_n+a_n}{2n^2}\alpha^2 \partial_\alpha^2 q(\alpha|x).
\end{equation}
In the limit we are looking for, all coefficients depend on $\epsilon=\alpha/n$, so we may write $a_n=a_n(\alpha/n)$, $b_n=b_n(\alpha/n)$, $c_n=c_n(\alpha/n)$. What is left is therefore to obtain sufficiently precise asymptotic estimates of those coefficients. Before proceeding to do this, we need two last preliminary steps, which are explained in the next subsections.

\subsection[\qquad\qquad\quad The regularizing matrices]{The regularizing matrices}\label{app:regularization}
We study here the matrix $E(\epsilon)=(E_{ij})_{0\leq i,j\leq n-1}$  with elements
\begin{equation}\label{eq:Ematrix}
    E_{ij}=\braket{x^{i+j}}=\frac{(i+j)!}{\epsilon^{i+j+1}},
\end{equation}
where the average is taken 
with respect to the Laguerre weight $x\mapsto e^{-\epsilon x}$ on $\mathbb{R}_+$. Similarly, we introduce $E'(\epsilon)=(E'_{ij})_{0\leq i,j\leq n-1}$  with elements
\begin{equation}\label{eq:Epmatrix}
    E'_{ij}=\braket{x^{i+j+\delta_{j,n-1}}}=\frac{(i+j+\delta_{j,n-1})!}{\epsilon^{i+j+1+\delta_{j,n-1}}}.
\end{equation}
Using the relation to orthogonal polynomials, the determinant and inverses of both matrices can be computed. We have
\begin{equation}
    \det E(\epsilon)=\frac{\prod_{k=0}^{n-1} k!^2}{\epsilon^{n^2}},
\end{equation}
and 
\begin{equation}
    E^{-1}_{ij}=\left.\frac{\partial^{i+j}K_n^{(\epsilon)}(x,y)}{i!j!\partial x^i \partial y^i}\right|_{x=y=0},
\end{equation}
with Chritoffel-Darboux kernel
\begin{equation}\label{eq:cd_laguerre}
    K_n^{(\epsilon)}(x,y)=n\frac{L_{n-1}(\epsilon x)L_{n}(\epsilon y)-L_n(\epsilon x)L_{n-1}(\epsilon y)}{x-y}.
\end{equation}
Using the explicit expressions for the Laguerre polynomials, one can simplify the inverse further, and obtain
\begin{equation}\label{eq:explicitinverse}
    E^{-1}_{ij}=\frac{(-1)^{i+j}\epsilon^{i+j+1}}{i!j!}\sum_{k=\max(i,j)}^{N-1} C_k^i C_k^j.
\end{equation}
The modified matrix $E'(\epsilon)$ can be treated as follows. First, notice that it differs from $E(\epsilon)$ only in the last column. This implies
\begin{equation}
    E^{-1}E'=\left(
    \begin{array}{ccccc}
    1&0&0&\ldots&u_0\\
    0&1&0&\ldots &u_1\\
    \vdots&&\ddots&&\vdots\\
    0&0&\ldots &1&u_{n-2}\\
    0&0&\ldots&0& u_{n-1}
    \end{array}
    \right)
    \qquad,\qquad u_j=\frac{(-1)^{n-j+1}}{\epsilon^{n-j}}C_n^j \frac{n!}{j!}.
\end{equation}
Therefore,
\begin{equation}
    \det E'(\epsilon)=\frac{n^2}{\epsilon}\det E(\epsilon).
\end{equation}
The matrix $E^{-1}E'$ can easily be inverted, which means
\begin{equation}
    E'^{-1}=\left(
    \begin{array}{ccccc}
    1&0&0&\ldots&v_0\\
    0&1&0&\ldots &v_1\\
    \vdots&&\ddots&&\vdots\\
    0&0&\ldots &1&v_{n-2}\\
    0&0&\ldots&0& n^2/\epsilon
    \end{array}
    \right) E^{-1}
    \qquad,\qquad v_i=\frac{(-1)^{n-i}}{\epsilon^{n-i-1}}\frac{C_n^i (n-1)!}{n i!}.
\end{equation}
This leads to the matrix elements
\begin{align}
    E'^{-1}_{ij}&=E^{-1}_{ij}+v_i E^{-1}_{n-1,j}\\\label{eq:modinverse}
    &=E^{-1}_{ij}-\frac{1}{n^2}(-\epsilon)^{i+j}\frac{(j+1)C_n^i C_{n}^{j+1}}{i! j!},
\end{align}
for $i\in\{0,\ldots,n-2\}$, and
\begin{equation}
    E'^{-1}_{n-1,j}=\frac{\epsilon}{n^2}E^{-1}_{n-1,j},
\end{equation}
for the inverse of $E'(\epsilon)$.

\subsection[\qquad\qquad\quad The Fredholm determinant]{The Fredholm determinant $\mathcal{Y}(\alpha)$}
\label{app:ortho_derivation2}
Consider the $n\times n$ matrix $M(\epsilon)=(M_{ij})_{0\leq i,j\leq n-1}$ with matrix elements 
\begin{equation}
    M_{ij}=\braket{x^{i+j}}
\end{equation}
with respect to the weight $w_\epsilon$, and introduce
\begin{equation}\label{eq:yndef}
    Y_n (\epsilon)=\frac{\det M(\epsilon)}{\det E(\epsilon)}.
\end{equation}
Our aim is to compute
\begin{equation}\label{eq:Ddef}
    \mathcal{Y}(\alpha)=\lim_{n\to\infty} Y_n(\alpha/n)
\end{equation}
We can do this using the method of Ref.\cite{Slavnov} which relies on the inversion formula (\ref{eq:inverse}). Write \begin{align}
    \det(E^{-1}(\epsilon)M(\epsilon))&=\det(I_n-E^{-1}(\epsilon)[E(\epsilon)-M(\epsilon)])
\end{align}
Now both $E^{-1}(\alpha/n)$ and the difference $E(\alpha/n)-M(\alpha/n)$ are well defined in the limit $n\to\infty$. The relevant polynomials for the inverse are the Laguerre polynomials; as explained in \ref{app:examples} those scale to Bessel functions. We obtain
\begin{align}
    \mathcal{Y}(\alpha)&=\det_{0\leq i,j\leq \infty}\left(\delta_{ij}-\sum_{k=0}^{\infty}
    \left.\frac{\partial^{i+k}K(x,y)}{i!k!\partial x^i \partial y^k}\right|_{x=y=0}\int_{\mathbb{R}}dz\,z^{k+j}\left[\Theta(z)-w(z)\right]\right)\\
    &=\det_{0\leq i,j\leq \infty}\left(\delta_{ij}-\sum_{k=0}^{\infty}
    \int_{\mathbb{R}}dz\,\left.\frac{\partial^i K(x,z)}{i!\partial x^i }\right|_{x=0}z^{k+j}\left[\Theta(z)-w(z)\right]\right)\\
    &=\det_{0\leq i,j\leq \infty}\left(\delta_{ij}-\left.\frac{\partial^i}{i!\partial x^i}\int_{\mathbb{R}}dz\, z^j V(x,z)\right|_{x=0}\right)
\end{align}
Here $K$ is defined as the limit
\begin{align}
    K(x,y)&=\lim_{n\to\infty} K_n^{(\alpha/n)}(x,y)\\
    &=\frac{\sqrt{\alpha y}J_0(2\sqrt{\alpha x}) J'_0(2\sqrt{\alpha y})-\sqrt{\alpha x}J_0(2\sqrt{\alpha y}) J_0'(2\sqrt{\alpha x})}{x-y}
\end{align}
where recall $K_n^{(\epsilon)}$ is the Christoffel Darboux kernel (\ref{eq:cd_laguerre}) corresponding to the Laguerre polynomials. We have also used the notation
\begin{equation}\label{eq:kernelbis}
    V(x,y)=K(x,y)\left[\Theta(y)-w_0(y)\right].
\end{equation}
This may be rewritten as the Fredholm determinant
\begin{equation}\label{eq:main_fredholm}
    \mathcal{Y}(\alpha)=\det(I-V)
\end{equation} with kernel (\ref{eq:kernelbis}), by writing a trace expansion and exchanging summations and integrals. This result can be used to compute the partition function $Z(\tau)$ defined by (\ref{eq:Z}). As explained in section \ref{sec:derivation}, it is given by the limit
\begin{equation}
    Z\left(\frac{\tau}{\sin\gamma}\right)=\lim_{n\to\infty} \frac{\left[\sin (\tau/n)\right]^{n^2}}{\prod_{k=0}^{n-1}k!^2} \det M(\tau/n)
\end{equation}
Using our previous formulae, we obtain
\begin{align}
    Z\left(\frac{\tau}{\sin\gamma}\right)&=\lim_{n\to\infty}  \left[\frac{n}{\tau}\sin \frac{\tau}{n}\right]^{n^2} Y_n(\tau/n)\\
    &=e^{-\tau^2/6}\mathcal{Y}(\tau)
\end{align}
corresponding to the choice of weight (\ref{eq:example_6v}) with $\cos\gamma=\Delta$.
This reproduces the result of Ref.~\cite{Stephan_return}, or  (\ref{eq:Z_theexact},\ref{eq:kern_main}) after a simple rescaling.

\subsection[\qquad\qquad\quad Asymptotics of the recurrence coefficients]{Asymptotics of the recurrence coefficients}
\label{app:ortho_derivation3}
Let us start with the two coefficients $a_n(\alpha/n)$ and $c_n(\alpha/n)$, which are easiest. Using (\ref{eq:an}), (\ref{eq:cn}), (\ref{eq:yndef}), we obtain
\begin{equation}\label{eq:cnbis}
    c_n(\alpha/n)=\frac{n^2}{\alpha} \frac{\sqrt{Y_{n-1}(\alpha/n)Y_{n+1}(\alpha/n)}}{Y_n(\alpha/n)}.
\end{equation}
Now we once again assume the form $Y_n(\epsilon)=\sum_{k=0}^{n^2}y_k n^k \epsilon^k$ holds for large $n$, small $\epsilon$. This implies $\mathcal{Y}(\alpha)=\lim_{n\to\infty} Y_n(\alpha/n)=\sum_{k\geq 0}y_k\alpha^k$, and
\begin{equation}
    Y_{n+p}(\alpha/n)-Y_n(\alpha/n)=\frac{p}{n}\alpha\partial_\alpha \mathcal{Y}(\alpha)+\frac{p^2}{2n^2}\alpha^2\partial_\alpha^2 \mathcal{Y}(\alpha)+o(1/n^2).
\end{equation}
Plugging this expansion in (\ref{eq:cnbis}) yields 
\begin{equation}\label{eq:cn_as}
    c_n(\alpha/n)=\frac{n^2}{\alpha}+\frac{\alpha}{2}\frac{d^2\log \mathcal{Y}(\alpha)}{d\alpha^2}+o(1).
\end{equation}
By a similar calculation
\begin{equation}\label{eq:an_as}
    a_n(\alpha/n)=\frac{n^2}{\alpha}+\frac{n}{\alpha}+\frac{\alpha}{2}\frac{d^2\log \mathcal{Y}(\alpha)}{d\alpha^2}+o(1).
\end{equation}
The central coefficient is much more complicated. Recall
\begin{equation}
    b_n(\epsilon)=\frac{D'_{n+1}(\epsilon)}{D_{n+1}(\epsilon)}-\frac{D'_{n}(\epsilon)}{D_{n}(\epsilon)}.
\end{equation}
Using the matrices $E(\epsilon), E'(\epsilon)$ introduced in \ref{app:regularization}, we may write
\begin{equation}
    \frac{D'_n(\epsilon)}{D_n(\epsilon)}=\frac{n^2}{\epsilon}R_n(\epsilon),
\end{equation}
where
\begin{equation}\label{eq:rn_def}
    R_n(\epsilon)=\frac{\det (E'^{-1}M')}{\det (E^{-1}M)}.
\end{equation}
Assume for now the following expansion holds
\begin{align}\label{eq:Rdef}
    R_n(\alpha/n)&=1-\frac{\mathcal{R}(\alpha)}{n^2}+O(n^{-3}),
\end{align}
[This will be justified later.] 
For consistency reasons
\begin{align}
    R_{n+1}(\alpha/(n+1))&=1-\frac{\mathcal{R}(\alpha)}{(n+1)^2}+O((n+1)^{-3})\\
    &=R_n(\alpha/n)+\frac{2\mathcal{R}(\alpha)}{n^3}+o(n^{-3}),
\end{align}
and
\begin{align}
    R_{n+1}(\alpha/n-\alpha/n^2)
    =R_{n+1}(\alpha/n)+\frac{\alpha\partial_\alpha \mathcal{R}(\alpha)}{n^3}+o(n^{-3}).
\end{align}
Equating the two expansions yields
\begin{equation}
    R_{n+1}(\alpha/n)-R_n(\alpha/n)=\frac{2\mathcal{R}(\alpha)-\alpha\partial_\alpha \mathcal{R}(\alpha)}{n^3}+o(n^{-3}),
\end{equation}
so it holds
\begin{align}
    b_n(\alpha/n)&=\frac{n}{\alpha}\left[(n+1)^2R_{n+1}(\alpha/n)-n^2R_n(\alpha/n)\right]\\
    &=\frac{2n^2}{\alpha}+\frac{n}{\alpha} -\partial_\alpha \mathcal{R}(\alpha)+o(n^0).
\end{align}
What is left is the computation of $\mathcal{R}(\alpha)$, which is defined as the limit
\begin{equation}
    \lim_{n\to\infty} n^2\left[1-R_n(\alpha/n)\right]
\end{equation}
where $R_n(\alpha/n)$ is given by (\ref{eq:rn_def}). Rewrite this as
\begin{equation}\label{eq:dettocompute}
    R_n=\det(I+(E^{-1}M)^{-1} B),
\end{equation}
where
\begin{align}
B&=E'^{-1}M'-E^{-1}M\\
    &=E'^{-1}(M'-E')-E^{-1}(M-E)\\\label{eq:Abis}
    &=\left(E'^{-1}-E^{-1}\right)(M-E)+E'^{-1}(\delta M-\delta E),
\end{align}
where we defined $\delta E=E'-E$, and $\delta M=M'-M$, matrices which consists of zeros except for the last column.
Now we are ready to take the Hamiltonian limit. Using (\ref{eq:modinverse}), one gets
\begin{equation}\label{eq:tricky}
    E'^{-1}(\alpha/n)-E^{-1}(\alpha/n)\sim -\frac{1}{n^2}\partial_\alpha E^{-1}(\alpha/n).
\end{equation}
In addition to that $M(\epsilon)-E(\epsilon)$ is regular as $\epsilon\to 0$, and the last term that involves $\delta E$, $\delta M$ on the rhs of (\ref{eq:Abis}) is negligible. The perturbation to the identity in the determinant (\ref{eq:dettocompute}) is very small; therefore, we may approximate
\begin{align}
    R_n&\simeq \det\left(I+(E^{-1}M)^{-1} (E'^{-1}-E^{-1})(M-E)\right)\\
    &\simeq 1+\textrm{Tr}\left[(E^{-1}M)^{-1}(E'^{-1}-E^{-1})(M-E)\right]\\
    &\simeq 1+\frac{1}{n^2}\textrm{Tr}\left[\left\{I-E^{-1}(\alpha/n)(E(0)-M(0))\right\}^{-1}\alpha\partial_\alpha E^{-1}(\alpha/n)(E(0)-M(0))\right]
\end{align}
Now \cite{Slavnov}, writing $E^{-1}M=I-E^{-1}(E-M)$ yields $\textrm{Tr}(E^{-1}(\alpha/n)M(\alpha/n))\to \textrm{tr}(I-V)$ where $V$ is the kernel (\ref{eq:ortho_kernel}). In a similar fashion, we get
\begin{equation}
    R_n(\alpha/n)= 1+\frac{1}{n^2} \textrm{tr}\left[ (I-V)^{-1}\alpha\partial_\alpha V\right]+o(1/n^2),
\end{equation}
which implies
\begin{align}
    \mathcal{R}(\alpha)
    &=-\textrm{tr}\left[ (I-V)^{-1}\alpha\partial_\alpha V\right]\\
    &=\alpha\partial_\alpha \textrm{tr} \log(I-V),
\end{align}
which may be rewritten as 
\begin{equation}\label{eq:rfinal}
    \mathcal{R}(\alpha)=\alpha\partial_\alpha \log\mathcal{Y}(\alpha).
\end{equation}
Summing everything up, the asymptotics of the diagonal coefficient read
\begin{equation}\label{eq:bn_as}
    b_n(\alpha/n)=\frac{2n^2}{\alpha}+\frac{n}{\alpha}-\partial_\alpha\left(\alpha\partial_\alpha \log \mathcal{Y}\right)+o(n^0),
\end{equation}
and we are in a position to justify the main claim of this appendix. Using (\ref{eq:cn_as},\ref{eq:an_as},\ref{eq:bn_as}) we obtain
\begin{align}
    \lim_{n\to\infty}\left[ b_n(\alpha/n)-a_n(\alpha/n)-c_n(\alpha/n)\right]&=-(2\alpha\partial_\alpha^2+\partial_\alpha)\log \mathcal{Y}\\
    \lim_{n\to \infty} \frac{a_n(\alpha/n)-c_n(\alpha/n)}{n}&=\frac{1}{\alpha}\\
    \lim_{n\to \infty} \frac{a_n(\alpha/n)+c_n(\alpha/n)}{2n^2}&=\frac{1}{\alpha}
\end{align}
and inserting those in (\ref{eq:crucial_step}) yields the ODE (\ref{eq:ortho_ode}).

\pagebreak
\section[\qquad\qquad\quad Power series checks]{Power series checks}
\label{app:checks}

This appendix is devoted to numerical checks of our main exact formulae (\ref{eq:pde}),(\ref{eq:Y}),(\ref{eq:multiamplitude}). The main idea is to expand the evolution operator $e^{\tau H}$ in power series, and exploit the fact that the action of $H^n$ on the initial state generates only a finite number of states. For example, the first few terms are
\begin{equation}
    e^{\tau H}\ket{\psi}=\ket{\psi}+\tau\left(-\Delta\ket{\psi}+\ket{\psi_1}\right)+\frac{\tau^2}{2}\left([1+\Delta^2]\ket{\psi}-4\Delta\ket{\psi_1}+\ket{\psi_2}+\ket{\psi_{01}}\right)+O(\tau^3)
\end{equation}
where we have used the notation (\ref{eq:psi_notation}) of the introduction. Higher orders become cumbersome quite quickly, but computer generation of those to high order can be done. Using standard bijections between fermionic configurations and Young diagrams \cite{Kac}, the number of states one needs to keep up to order $n$ is the number of Young diagrams of size at most $n$. Those can be generated using known routines, and their number is the partition number $p(n)$, which by the Hardy-Ramanujan formula grows exponentially fast with $\sqrt{n}$.

We were able to compute simple observables up to order $\approx 70$ in $\tau$, for any $\Delta$, using this brute force method\footnote{Specifying a given values of $\Delta$, e.g. $\Delta=3/4$ allows to reach even higher order.}. For example the partition function is given by
\begin{align}\nonumber
   Z(\tau)&=1-\Delta  \tau +\frac{1}{2} \left(\Delta ^2+1\right) \tau ^2-\frac{1}{6} \Delta  \left(\Delta ^2+5\right) \tau ^3+\frac{1}{24} \left(\Delta ^4+18 \Delta ^2+3\right) \tau ^4\\&-\frac{1}{120} \Delta  \left(\Delta ^4+58 \Delta ^2+31\right) \tau ^5+\frac{1}{720} \left(\Delta ^6+179 \Delta ^4+201 \Delta ^2+15\right) \tau ^6+O(\tau^7)
   \label{eq:powerseries_Z}
\end{align}
where we do not write the higher order terms for obvious space reasons. 
This result can be compared to a power series expansion of the partition function  
\begin{equation}\label{eq:ZtoD}
    Z(\tau)=e^{-\tau^2\sqrt{1-\Delta^2}/6}\mathcal{Y}(\tau\sqrt{1-\Delta^2})
\end{equation} where the Fredholm determinant $\mathcal{Y}$ (see (\ref{eq:main_fredholm})) can be rewritten as \cite{Stephan_return}
\begin{equation}\label{eq:infdet}
    \mathcal{Y}(\alpha)=\det_{0\leq i,j\leq \infty}\left(\delta_{ij}-\sum_{k=0}^\infty \frac{\Gamma_{k+j+1}(\tau)}{j!^2k!^2(i+k+1)}\right)
\end{equation}
where 
\begin{equation}
    \Gamma_q(\tau)=\left(\frac{\tau}{\pi}\right)^q (q-1)! \left[((-1)^q+1)\zeta(q)+(-1)^{q+1}\zeta(q,\gamma/\pi)-\zeta(q,1-\gamma/\pi)\right].
\end{equation}
 $\zeta(q,g)=\sum_{n\geq 0}\frac{1}{(n+g)^q}$ is the Hurwitz zeta function, while $\zeta(q)=\zeta(q,1)$ is the Riemann zeta function. The case $q=1$ is understood as a limit, $\Gamma_1(\tau)=\lim_{q\to 1}\Gamma_q(\tau)=\tau \cot \gamma$.  Since $\Gamma_q$ is proportional to the monomial $\tau^q$, one can easily get a series expansion of $\mathcal{Y}(\tau)$ up to order $p$ by truncating (\ref{eq:infdet}) to some finite $p\times p$ determinant, and truncating the sum over $k$ in (\ref{eq:infdet}) accordingly. This yields
 \begin{align}\nonumber
    &\mathcal{Y}(\tau)=1-\cos (\gamma )\left(\frac{\tau}{\sin\gamma}\right)+\frac{1}{6}  (\cos 2 \gamma +5)\left(\frac{\tau}{\sin\gamma}\right)^2-\cos \gamma \left(\frac{\tau}{\sin\gamma}\right)^3+\\
    &\frac{1}{36} \left(21+12 \cos 2 \gamma -\sin ^4\gamma \right)\left(\frac{\tau}{\sin\gamma}\right)^4+\frac{\cos \gamma}{180} \left(\sin ^4\gamma +60 \sin ^2\gamma -135\right) \left(\frac{\tau}{\sin\gamma}\right)^5 +O(\tau^6),
\end{align}
From this one can recover $Z(\tau)$ using (\ref{eq:ZtoD}), and we get perfect agreement with the expression (\ref{eq:powerseries_Z}). Unfortunately we were only able to perform the check up to order $12$ in that case, because the Taylor expansion of the determinant (\ref{eq:infdet}) is extremely heavy to compute. For this reason, we rather use the Young diagram method to compute $Z(\tau)$ in the remainder of the appendix. 
One easily obtains $Q(\tau)$ using (\ref{eq:Y}): 
\begin{align}\nonumber
Q(\tau)&=-\Delta +3 \tau-5 \Delta  \tau ^2+\frac{14 \Delta ^2 \tau ^3}{3}-\frac{3}{2} \Delta  \left(2 \Delta ^2-1\right) \tau ^4+
    \frac{11}{15} \Delta ^2 \left(2 \Delta ^2-5\right) \tau ^5\\
&-\frac{1}{180} 13 \Delta  \left(8 \Delta ^4-64 \Delta ^2+5\right) \tau ^6+\frac{1}{42} \Delta ^2 \left(8 \Delta ^4-168 \Delta ^2+63\right) \tau ^7+O(\tau^8)
\label{eq:series_Q}
\end{align}
We are now in a position to check the PDE (\ref{eq:pde}) for $h(\tau|z)$, since $Q(\tau)$ was the only unknown term left in it. Using the nearest neighbor locality of $H$, we obtain the expansion
\begin{equation}
    h(\tau|z)=\sum_{m\neq 0} q_m(z)\tau^m 
\end{equation}
where $q_m$ is a polynomial of degree $m$ in $z$,
and solve the PDE order by order. The first few orders read
\begin{align}\nonumber
    h(\tau|z)&=1+z\tau+z(z-2\Delta)\frac{\tau^2}{2!}+z(z-2\Delta)^2\frac{\tau^3}{3!}+z\left[(z-2\Delta)^3+4\Delta\right]\frac{\tau^4}{4!}\\
    &+z\left[(z-2\Delta)^4+10\Delta(z-4\Delta) \right]\frac{\tau^5}{5!}+O(\tau^6)
    \label{eq:series_h}
\end{align}
and one can check to high orders a perfect match with the alternative expression for $h(\tau|z)$ obtained from the brute force method.

It is also possible to make several other consistency checks of our results. For example the Hamiltonian limit of the orthogonal polynomials (see (\ref{eq:ortholimit})) which play a key role in our derivation can be obtained from the brute force method, or from the ODE (\ref{eq:ortho_ode}) with a source term once again related to the partition function. Writing $q(\tau|x)=\sum_{k\geq 0} r_k(x)\tau^k$, the first few $r_k$ read
\begin{align}
    r_0(x)&=1\\
    r_1(x)&=\frac{\Delta }{\sqrt{1-\Delta ^2}}-x\\
    r_2(x)&=-\frac{\Delta ^2-2}{2 \left(\Delta ^2-1\right)}+\frac{x^2}{4}-\frac{\Delta  x}{2 \sqrt{1-\Delta ^2}}\\
    r_3(x)&=\frac{\Delta ^3}{6 \left(1-\Delta ^2\right)^{3/2}}-\frac{x^3}{36}+\frac{\Delta  x^2}{12 \sqrt{1-\Delta ^2}}+\frac{\left(5-2 \Delta ^2\right) x}{9-9 \Delta ^2}
\end{align}
with perfect match for higher order ones. 

Another example is the exact formula for the EFP for small values of $l$. At $l=1$ it is easy to see that the contour integral (\ref{eq:efp_delta_imag}) simply selects the coefficient of $z^0$ in the product $h(\frac{R-y}{2}|z)h(\frac{R+y}{2}|1/z)$. The lowest order expansion is
\begin{align}
   \frac{\mathcal{F}_1}{\mathcal{F}_0} &=1+\frac{R^2}{4} \left(1-Y^2\right)-\frac{R^3}{4} \Delta   \left(1-Y^2\right)\\\nonumber
    &\frac{R^4}{192} \left(1-Y^2\right) \left(28 \Delta ^2+\left(4 \Delta ^2-3\right) Y^2+3\right)
    -\frac{R^5}{96} \Delta  \left(1-Y^2\right) \left(6 \Delta ^2+\left(2 \Delta ^2-5\right) Y^2+1\right)+\ldots
\end{align}
where $Y=y/R$ is set to some fixed value in $(-1,1)$.

For $l=2$ one needs the two-particle analog of $h(\tau|z)$ (see (\ref{eq:final_multih})) which has series expansion 
\begin{align}\nonumber
h(\tau|z_1,z_2)&=1+z_1 z_2+\tau  \left[(1+z_1 z_2)(z_1+z_2)-2\Delta z_1 z_2\right]\\
&+\frac{\tau^2}{2}\left[(z_1+z_2)^2(1+z_1 z_2)-2\Delta (z_1+z_2)(1+2z_1 z_2)+4\Delta^2 z_1 z_2\right]+O(\tau^3).
\end{align}
Now, since we are integrating on a small contour centered around the origin, one can use the following series expansion 
\begin{equation}
    \prod_{1\leq i\neq j\leq 2} \frac{z_k-z_j}{z_j z_k-2\Delta z_j+1}=-(z_1-z_2)^2\sum_{p,q\geq 0}(-1)^{p+q}(z_j z_k-2\Delta z_j)^p(z_j z_k-2\Delta z_k)^q
\end{equation}
and the contour integral formula just selects (half) the coefficient of $z_1^0 z_2^0$ in the product of the rhs right above with $\tH(\frac{R-y}{2},\frac{1}{z_1},\frac{1}{z_2})\tH(\frac{R+y}{2},z_1,z_2)$. We show the result at $y=0$ for simplicity, which is given by
\begin{equation}
    \frac{\mathcal{F}_2}{\mathcal{F}_0}=1+\frac{R^2}{4}-\frac{\Delta  R^3}{4}+\frac{1}{96} \left(14 \Delta ^2+3\right) R^4
    -\frac{1}{32} \Delta  \left(2 \Delta ^2+1\right) R^5+O(R^6).
\end{equation}
which matches the results obtained from the brute force method to high orders.

\pagebreak
\section*{References}

\bibliographystyle{iopart-num}
\bibliography{biblio}
\end{document}